\documentclass[12pt,a4paper]{article}
\usepackage[top=30truemm,bottom=30truemm,left=20truemm,right=20truemm]{geometry}
\usepackage[dvipdfmx]{graphicx}
\usepackage{atbegshi}
\usepackage{amsmath} 
\usepackage{amssymb}
\usepackage{amsfonts}
\usepackage{authblk}
\usepackage{fancyhdr}
\usepackage{comment} 
\usepackage{multicol} 
\usepackage{titlesec} 
\usepackage{ifthen}

\usepackage{color}

\usepackage{secdot}
  \sectiondot{subsection}

\usepackage{lastpage} 
\makeatletter 
    
    \@addtoreset{equation}{section}
  \makeatother
  
\begin{document}
\thispagestyle{fancy}
  \rhead{\ifthenelse{\value{page}=1}{RUP-18-4}{\thepage{}}}
  
\renewcommand{\thefootnote}{\fnsymbol{footnote}} 
\vspace*{0.7cm}
\begin{center}
\large{\textbf{Three ways to solve critical $\phi^4$ theory on $4-\epsilon$ dimensional real projective space: perturbation, bootstrap, and Schwinger-Dyson equation}}
\end{center}
\vspace*{1.5cm}
\begin{center}
\normalsize{Chika Hasegawa$^{1}$ \footnote[1]{\texttt{chika.hasegawa AT rikkyo.ac.jp}}
and Yu Nakayama$^{1}$  \footnote[2]{\texttt{yu.nakayama AT rikkyo.ac.jp}}}  \\
\vspace*{1.0cm}
${}^1$\textit{Department of Physics, Rikkyo University, Tokyo 171-8501, Japan} \\
\end{center}
\vspace*{3.8cm}
\begin{abstract}

We solve the two-point function of the lowest dimensional scalar operator in the critical  $\phi^4$ theory on $4-\epsilon$ dimensional real projective space in three different methods. The first is to use the conventional perturbation theory, and the second is to impose the crosscap bootstrap equation, and the third is to solve the Schwinger-Dyson equation under the assumption of conformal invariance. We find that the three methods lead to mutually consistent results but each has its own advantage.

\end{abstract}

\newpage

\renewcommand{\thefootnote}{\arabic{footnote}} 
\section{Introduction} \label{Introduction}
The legend says that Feynman claims ``I feel I really understand something when I can give more than two explanations."\footnote{This is actually unsourced. One  can find the quote in Japanese Wikipedia article on Feynman as of January 2018 with no further reference.} In theoretical physics it is sometimes crucial to derive the same results from different methods: unlike mathematics, we do not really have a proof of the consistency of interacting quantum field theories as we know today, so it is a priori unclear if different approaches to the same problem should give the same answer. Take a quantum anomaly, for instance; it took quite a while to recognize it is really the actual phenomenon rather than the mere failure of particular computational methods: it is only after we got convinced that every conceivable methods give the same answer. Moreover, since equivalence of different methods are non-obvious, one may obtain novel mathematical identities or hints for further non-trivial dualities. This is probably what Feynman wanted, and it is how and why he invented the path integral.

Conformal field theories (CFTs) have many applications in physics, so in order to deepen our understanding (especially in Feynman's way), it is imperative to establish the mutual consistency of different ways to solve them. In particular, the recent revival \cite{Rattazzi:2008pe} of conformal bootstrap approach \cite{Ferrara:1973yt}\cite{Polyakov:1974gs}\cite{Mack:1975jr} revealed the power of conformal symmetry even without using the explicit Hamiltonian or Lagrangian.  It is based on abstract operator algebra together with the constraint from the crossing symmetry of four-point functions. The result is surprising: it numerically solved the three-dimensional critical Ising model \cite{ElShowk:2012ht}\cite{El-Showk:2014dwa}\cite{Kos:2014bka}\cite{Kos:2016ysd} to the precision the other methods had never reached. The question, however, remained: how can we solve the critical Ising model without knowing that we are actually solving the critical Ising model?

Conventional studies of quantum field theories are based on Hamiltonian or Lagrangian. One may perform perturbative expansions and compute Feynman diagrams. Beyond perturbation theories, one may try to extract the full information of Hamiltonian or Lagrangian by solving the Schwinger-Dyson equations. In principle, this gives a non-perturbative approach to the quantum field theory under consideration while we may have to face various questions such as the non-perturbative renormalization or a choice of ``correct" solutions of the Schwinger-Dyson equation. Even if we find a particular solution for a particular equation, we do not know if these solutions are consistent as an entire theory.

In recent years, there have been consistency checks of these different approaches in the perturbative regime in the critical $\phi^n$ Landau-Ginzburg models on flat Euclidean space-time. Whenever  the comparison can be made in the perturbative regime, the prediction from the conformal bootstrap\footnote{In \cite{Gopakumar:2016wkt}\cite{Gopakumar:2016cpb}\cite{Dey:2016mcs} (see also \cite{Dey:2017fab}\cite{Dey:2017oim}), they have developed analytic approaches to the conformal bootstrap in perturbative regimes by using the Mellin space. On the other hand, in \cite{Alday:2015eya}\cite{Alday:2015ewa}\cite{Alday:2016njk}\cite{Alday:2016jfr}, they have developed a large spin perturbation theory as another analytic bootstrap approach, and obtained the CFT data at the Wilson-Fisher fixed point to quartic order in $\epsilon$ \cite{Alday:2017zzv}.} agrees with the ones from the Schwinger-Dyson approach \cite{Nii:2016lpa} or more sophisticated refinement of this approach known as ``$\epsilon$-expansions in conformal field theories"\cite{Rychkov:2015naa} (see also analysis of various models \cite{Basu:2015gpa}\cite{Ghosh:2015opa}\cite{Raju:2015fza}, with defects \cite{Yamaguchi:2016pbj}, or with spinning operators \cite{Roumpedakis:2016qcg}). The gist is that under some assumptions, the CFT data of the critical Ising model (or $\phi^n$ Landau-Ginzburg models more generally) is the only available consistent set at least in the perturbative regime. See also \cite{Skvortsov:2015pea}\cite{Giombi:2016hkj}\cite{Roumpedakis:2016qcg}\cite{Liendo:2017wsn}\cite{Giombi:2017rhm} further in this direction.

Solving quantum field theories on non-trivial curved background is a challenging task. It is therefore an interesting question to address if we can use the conformal symmetry and non-perturbative techniques discussed above to solve CFTs on non-trivial curved background as in the flat space-time. Obviously, we may trivially solve CFTs on conformally flat manifold, in which all the conformal symmetry is preserved, by just rescaling all the correlation functions up to possible conformal anomaly. 
Our target in this paper, however, is a real projective space, which is locally conformally flat, but not globally.\footnote{See \cite{Nakayama:2016cim}\cite{Hasegawa:2016piv} for applications of CFTs on real projective space to critical phenomena, and \cite{Nakayama:2015mva}\cite{Nakayama:2016xvw} for to the AdS/CFT correspondence.} It preserves half of the original conformal symmetries on flat space-time. The central question is if the methods useful in solving conformal field theories in flat space-time are still powerful enough to solve them on real projective space-time. If so, such methods may be worthwhile studying further in other more non-trivial space-time.
As a theoretical interest, we also ask the following question: assuming that we know all the CFT data on a flat space-time, how much can we determine the CFT data on a curved space-time including a real projective space?

In this paper, we propose three different methods to solve the two-point function of the lowest dimensional scalar operator in the  critical $\phi^4$ theory on $4-\epsilon$ dimensional real projective space. 
We will see that the three methods lead to mutually consistent results but each has its own advantage. In particular, two out of the three ways i.e. the crosscap bootstrap approach and the Schwinger-Dyson equations are candidates for the non-perturbative studies, so the agreement in the perturbative regime may be regarded as the theoretical ground that would ensure the validity in the non-perturbative regime.

The organization of this paper is as follows.
In section \ref{Conformal field theory on real projective space}, we review some basic facts about conformal field theories on the $d$-dimensional real projective space.
In section \ref{Conventional perturbation theory}, we derive the two-point function of the lowest dimensional scalar operator in the critical $\phi^4$ theory on $4-\epsilon$ dimensional real projective space from the conventional perturbation theory in the weak coupling regime.
In section \ref{Crosscap bootstrap in order epsilon}, we solve the crosscap bootstrap equation for the $\phi$-$\phi$ two-point function to the first non-trivial order in $\epsilon$.
In section \ref{Schwinger-Dyson equation approach}, we derive some CFT data appearing in the $\phi$-$\phi$ two-point function by using the conformal symmetry and the Schwinger-Dyson equations applied to the $\phi$-$\phi$ two-point function.
In section \ref{Conclusion}, we conclude with some discussions.
In appendix \ref{phi-phi-Box2 phi4 three point function}, we demonstrate vanishing of a certain operator product expansion (OPE) coefficient in the critical $\phi^4$ theory at $O(\epsilon)$.
In appendix \ref{Formulae for Gauss's hypergeometric function}, we collect some formulae for Gauss's hypergeometric function used in the main text. 
In appendix \ref{Laplacian acting two-point functions}, we summarize the calculation of the  Laplacian acting on the two-point functions.
In appendix \ref{Laplacian with y=0}, we rederive the action of the Laplacian on the two-point function in a slightly simplified manner.

\section{Conformal field theory on real projective space} \label{Conformal field theory on real projective space}

In this section, we review some basic facts about conformal field theories on a $d$-dimensional real projective space, based on \cite{Nakayama:2015mva}\cite{Nakayama:2016cim}\cite{Nakayama:2016xvw}. 
A $d$-dimensional real projective space $\mathbb{RP}^d$ is defined by identification of antipodal points on a $d$-dimensional sphere $\mathbb{S}^d$. To study CFTs on it, we may use the conformal mapping and  define a (conformally equivalent) $d$-dimensional real projective space $\mathbb{RP}^d$ by involution $\vec{x} \to - \frac{\vec{x}}{|\vec{x}|^2}$ for $d$-dimensional Cartesian coordinate vector $\vec{x} = (x^1,x^2,\cdots,x^d)$ on a $d$-dimensional flat Euclidean space $\mathbb{R}^d$.
In this paper we take the fundamental region of the $d$-dimensional real projective space $\mathbb{RP}^d$ as $0 \le |\vec{x}| \le 1$.
Identification of operators on each antipodal points breaks the Euclidean conformal symmetry $SO(d+1,1)$ down to its subgroup $SO(d+1)$. 
Invariance of the remaining conformal symmetry $SO(d+1)$ on $\mathbb{RP}^d$ fixes a functional form of correlation functions. 
For instance, one-point functions of scalar primary operators $O_i$ with conformal dimension $\Delta_{i}$ are determined as
\begin{align}
\langle O_i (\vec{x}) \rangle^{\mathbb{RP}^d} = \frac{A_i^{\Omega}}{(1+|\vec{x}|^2)^{\Delta_i}}, \label{eq:1pt func in CFT on RPd}
\end{align}
where $A_i^{\Omega}$ are additional CFT data on the real projective space. We note that the one-point functions $A_i^{\Omega}$ depend on the involution $\Omega$ to define the CFT on $\mathbb{RP}^d$. In our study of the critical $\phi^4$ theory, we may use the global $\mathbb{Z}_2$ symmetry in combination with the geometrical identification, and we have two choices of $\Omega = \pm$. Here $+$ corresponds to a trivial involution while $-$ corresponds to the involution in combination with the $\mathbb{Z}_2$ symmetry.
We also note that one-point functions of spinning operators vanish due to the $SO(d+1)$ invariance. Therefore, solving CFTs on $\mathbb{RP}^d$ is equivalent to specifying all the $A_i^{\Omega}$.

Similarly, two-point functions of each scalar primary operator $O_i$ with conformal dimension $\Delta_{i}$ are fixed up to a function of a single conformal invariant parameter  $\eta := \frac{|\vec{x}_1-\vec{x}_2|^2}{(1+|\vec{x}_1|^2)(1+|\vec{x}_2|^2)}$, which is called the crosscap crossratio:
\begin{align}
\langle O_i (\vec{x}_1) O_j (\vec{x}_2) \rangle^{\mathbb{RP}^d} = \frac{(1+|\vec{x}_1|^2)^{\frac{-\Delta_i + \Delta_j}{2}} (1+|\vec{x}_2|^2)^{\frac{-\Delta_j + \Delta_i}{2}} }{|\vec{x}_1-\vec{x}_2|^{2\left( \frac{\Delta_i + \Delta_j}{2} \right)} } G_{ij}^{\Omega}(\eta). \label{eq:2pt func in CFT on RPd}
\end{align}
Here, $G^\Omega_{ij}(\eta)$ depends on the theory and the choice of the involution $\Omega$.
From the locality of the CFT,  we can further decompose them by conformal partial waves as follows
\begin{align}
G_{ij}^{\Omega}(\eta) = \sum_{k} C_{ij}^{\ \ k} A_{k}^{\Omega} \eta^{\frac{\Delta_k}{2}} {}_2 F_1\left(\frac{ \Delta_i - \Delta_j + \Delta_k}{2},\frac{ \Delta_j - \Delta_i + \Delta_k}{2};\Delta_k + 1 - \frac{d}{2};\eta \right), \label{eq:CPWD in CFT on RPd}
\end{align}
where $C_{ij}^{\ \ k}$ are OPE coefficients (raised by the Zamolodchikov metric) and $A_{k}^{\Omega}$ are the one-point function coefficients which appeared above \eqref{eq:1pt func in CFT on RPd}. Since one-point functions vanish for spinning operators, the sum is taken only over the scalar primary operators in the theory.

Let us now review the concept of crossing symmetry on $\mathbb{RP}^d$. The definition of CFTs on $\mathbb{RP}^d$ makes us identify the operators on a point of $\mathbb{R}^d$  with that on its antipodal point up to a choice of the involution.
To compute the two-point function, we can either take the OPE as $\vec{x}_1$ to $\vec{x}_2$, or  we can take the OPE as $\vec{x}_1$ to $\tilde{\vec{x}}_2 = - \frac{\vec{x}_2}{|\vec{x}_2|^2}$. The identification under the involution demands they must be related:
\begin{align}
\left( \frac{1-\eta}{\eta^2} \right)^{\frac{\Delta_i + \Delta_j}{6}} G^{\Omega}_{ij}(\eta) = \Omega_{j}^{\ k} \left( \frac{\eta}{(1-\eta)^2} \right)^{\frac{\Delta_i + \Delta_j}{6}} G_{ik}^{\Omega}(1-\eta). \label{eq:crosscap bootstrap equation 0}
\end{align}
This crossing equation is known as the crosscap bootstrap equation. In our study of the critical $\phi^4$ theory, the choice of the involution $\Omega=\pm$ is based on the $\mathbb{Z}_2$ symmetry, so the matrix $\Omega_{j}^{\ k}$ is just the sign $\pm$ (depending on how it acts on the operator $O_k$).

\section{Conventional perturbation theory} \label{Conventional perturbation theory}

The first method we would like to pursue in this paper is the conventional perturbation theory. In this section, we study the two-point function of the lowest dimensional scalar operator in the critical $\phi^4$ theory on $4-\epsilon$ dimensional real projective space from the conventional perturbation theory in the weak coupling regime.

The classical action of the critical $\phi^4$ theory in $d=4-\epsilon$ dimensions is given by 
\begin{align}
S[\phi, g] = \int \mathrm{d}^d x \left[ \frac{1}{2}(\partial \phi)^2 + \frac{g}{4!} \phi^4 \right], \  \label{eq:action of critical phi4}
\end{align}
with respect to the elementary scalar field $\phi$.  The model is defined by the Euclidean path integral
\begin{align}
Z[g] = \int [\mathcal{D}\phi] \mathrm{e}^{-S[\phi, g]} \ ,
\end{align}
and we will evaluate it as a perturbative expansion with respect to the coupling constant $g$ around the Gaussian fixed point $g=0$. 

The theory on $\mathbb{RP}^d$ is defined by the involution $\Omega = \pm$ acting on the elementary field as $\phi(\vec{x}) \to \pm \phi(-\frac{\vec{x}}{|\vec{x}|^2}).$ At $g=0$, some of the free-field correlation functions on $\mathbb{RP}^d$ that we will use are obtained by the method of image as
\begin{align}
\langle \phi (\vec{x}) \rangle^{\mathbb{RP}^d}_{\mathrm{free}} &= 0, \label{eq:1pt func in 4-epsilon g=0 phi4 theory} \\
\langle \phi^2 (\vec{x}) \rangle^{\mathbb{RP}^d}_{\mathrm{free}} &= \pm \frac{1}{4\pi^2} \frac{1}{(1+|\vec{x}|^2)^{2\Delta_{\phi}^{\mathrm{free}}}}, \label{eq:phi2 correlation func in 4-epsilon g=0 phi4 theory} \\
\langle \phi (\vec{x}) \phi (\vec{y}) \rangle^{\mathbb{RP}^d}_{\mathrm{free}} &= \frac{1}{4\pi^2} \frac{1}{|\vec{x}-\vec{y}|^{2\Delta_{\phi}^{\mathrm{free}}}} \left[ 1 \pm \left( \frac{\eta}{1-\eta} \right)^{\Delta_{\phi}^{\mathrm{free}}} \right]. \label{eq:2pt func in 4-epsilon g=0 phi4 theory}
\end{align}
where $\frac{1}{4\pi^2}$ is a normalization factor\footnote{The normalization of $(4\pi^2)$ may not be a good one for the theory in $4-\epsilon$ dimension with finite $\epsilon$ because one may want to use the surface volume $S_d= 2\pi^{d/2}/\Gamma(d/2)$ times $(d-2)$ instead, but we opt to use this normalization. The choice does not affect any physical consequence, but we have to be careful about the normalization when we actually compare the amplitude such as $A_\phi^{\pm}$ or $C_{\phi\phi\phi}$ rather than the exponent.} (i.e. $C_{\phi \phi}^{\ \ I} A_{I}^{\pm}$), and $\Delta_{\phi}^{\mathrm{free}}=\frac{d-2}{2}$ is a conformal dimension of the elementary scalar field $\phi$ in the free theory. The signs here  and hereafter are correlated with the choice of the involution $\Omega$.

Using the perturbative expansions, we can evaluate the two-point function of $\phi$ via the Wick contraction as
\begin{align}
\langle \phi(\vec{x}) \phi(\vec{y}) \rangle^{\mathbb{RP}^d} &= \langle \phi(\vec{x}) \phi(\vec{y}) \rangle^{\mathbb{RP}^d}_{\mathrm{free}}
- \frac{g}{4!} \int \mathrm{d}^d z \langle \phi(\vec{x}) \phi(\vec{y}) \phi^4 (\vec{z}) \rangle^{\mathbb{RP}^d}_{\mathrm{free}} + O(g^2) \nonumber \\
&=\langle \phi(\vec{x}) \phi(\vec{y}) \rangle^{\mathbb{RP}^d}_{\mathrm{free}}
- \frac{g}{4!} \int \mathrm{d}^d z \left[ 3 \cdot \langle \phi(\vec{x}) \phi(\vec{y}) \rangle^{\mathbb{RP}^d}_{\mathrm{free}}  \left[ \langle \phi^2 (\vec{z}) \rangle^{\mathbb{RP}^d}_{\mathrm{free}} \right]^2 \right. \nonumber \\
& \quad \left. + 12 \cdot \langle \phi(\vec{x}) \phi(\vec{z}) \rangle^{\mathbb{RP}^d}_{\mathrm{free}} \langle \phi(\vec{y}) \phi(\vec{z}) \rangle^{\mathbb{RP}^d}_{\mathrm{free}}  \langle \phi^2 (\vec{z}) \rangle^{\mathbb{RP}^d}_{\mathrm{free}} \right] + O(g^2), \label{eq:2pt func in perturbarion theory}
\end{align}
where the integration domain is $0 \leq |z| \leq 1$. By an appropriate normalization, we will neglect the vacuum amplitude that appears in the first term at $O(g)$
and focus on the second term.
To facilitate the computation but without loss of generality,  by using $SO(d+1)$ symmetry, we may put the two points $\vec{x}$ and $\vec{y}$ on one straight line from the origin.\footnote{See \cite{McAvity:1993ue}\cite{McAvity:1995zd} for the studies of two-point functions in the the similar configurations in boundary CFTs.} We can further set $d=4$ in the $O(g)$ term in this approximation.

By substituting the free field correlation functions and doing explicit integration, we obtain the following expression for the first order perturbative correction to the two-point function:
\begin{align}
&\left(\langle \phi(\vec{x}) \phi(\vec{y}) \rangle^{\mathbb{RP}^d}\right)^{(1)}\epsilon :=- \frac{g}{4!} \int \mathrm{d}^d z \left[12 \cdot \langle \phi(\vec{x}) \phi(\vec{z}) \rangle^{\mathbb{RP}^d}_{\mathrm{free}} \langle \phi(\vec{y}) \phi(\vec{z}) \rangle^{\mathbb{RP}^d}_{\mathrm{free}}  \langle \phi^2 (\vec{z}) \rangle^{\mathbb{RP}^d}_{\mathrm{free}} \right] \nonumber \\
&\quad = \mp \frac{g}{2} \left( \frac{1}{4\pi^2} \right)^3 \int \mathrm{d}^4 z \left[ \frac{1}{|\vec{x}-\vec{z}|^2} \left[ 1 \pm \left( \frac{\eta_{xz}}{1-\eta_{xz}} \right) \right] \frac{1}{|\vec{y}-\vec{z}|^2} \left[ 1 \pm \left( \frac{\eta_{yz}}{1-\eta_{yz}} \right) \right] \frac{1}{(1+|\vec{z}|^2)^2} \right] \nonumber \\
&\quad = \mp \frac{g}{2} \cdot 4\pi  \left( \frac{1}{4\pi^2} \right)^3 \nonumber \\
&\qquad \times \int_{0}^{1} \mathrm{d} |\vec{z}| \frac{|\vec{z}|^3}{(1+|\vec{z}|^2)^2} \int_{0}^{\pi} \mathrm{d} \theta \sin^2 \theta \left[ \frac{1}{|\vec{x}|^2 - 2|\vec{x}||\vec{z}|\cos \theta + |\vec{z}|^2} \frac{1}{|\vec{y}|^2 - 2|\vec{y}||\vec{z}|\cos \theta + |\vec{y}|^2} \right. \nonumber \\ 
&\qquad \quad \pm \frac{1}{1+2 |\vec{x}||\vec{z}|\cos\theta + |\vec{x}|^2\vec{z}|^2} \frac{1}{|\vec{y}|^2 - 2|\vec{y}||\vec{z}|\cos \theta + |\vec{y}|^2} \nonumber \\
&\qquad \quad \pm \frac{1}{|\vec{x}|^2 - 2|\vec{x}||\vec{z}|\cos \theta + |\vec{z}|^2} \frac{1}{1+2 |\vec{y}||\vec{z}|\cos \theta + |\vec{y}|^2|\vec{z}|^2} \nonumber \\
&\qquad \quad \left. + \frac{1}{1+2 |\vec{x}||\vec{z}|\cos \theta + |\vec{x}|^2|\vec{z}|^2}  \frac{1}{1+2 |\vec{y}||\vec{z}|\cos \theta + |\vec{y}|^2|\vec{z}|^2} \right] \nonumber \\
&\quad = \frac{g}{2} \pi^2 \left( \frac{1}{4\pi^2} \right)^3 \left[ \pm \frac{1}{1 + 2 \vec{x}\cdot \vec{y} +|\vec{x}|^2 |\vec{y}|^2} \ln \frac{|\vec{x}-\vec{y}|^2}{(1+|\vec{x}|^2)(1+|\vec{y}|^2)} + \frac{1}{|\vec{x}-\vec{y}|^2} \ln \frac{1+2\vec{x}\cdot \vec{y} + |\vec{x}|^2 |\vec{y}|^2}{(1+|\vec{x}|^2)(1+|\vec{y}|^2)} \right]. \label{eq:2pt func in perturbarion theory order epsilon}
\end{align}
Here, we have defined various crosscap crossratios: $\eta_{xz} := \frac{|\vec{x}-\vec{z}|^2}{(1+|\vec{x}|^2)(1+|\vec{z}|^2)}$, $\eta_{yz} := \frac{|\vec{y}-\vec{z}|^2}{(1+|\vec{y}|^2)(1+|\vec{z}|^2)}$, and $\eta = \frac{|\vec{x}-\vec{y}|^2}{(1+|\vec{x}|^2)(1+|\vec{y}|^2)}$. Note that the assumption of the collinearity  gives $\vec{x}\cdot\vec{z} = |\vec{x}||\vec{z}| \cos\theta$ and $\vec{y}\cdot\vec{z} = |\vec{y}||\vec{z}| \cos\theta$ in the polar coordinate.

At this point, we have to impose the conformal invariance. To do this, we set the coupling constant $g$ to a critical value of the Wilson-Fisher fixed point so that the renormalization group beta function vanishes. From the one-loop beta function on $\mathbb{R}^d$, it is given by $g_{*} = \frac{16\pi^2}{3} \epsilon + O(\epsilon^2)$. In order to make the expression simpler, we further introduce the known anomalous dimension of $\phi^2$ operator (i.e. $\gamma_{\phi^2} = (\gamma_{\phi^2})^{(1)}\epsilon + O(\epsilon^2) = \frac{1}{3}\epsilon + O(\epsilon^2)$), and the normalization factor $C_{\phi \phi}^{\ \ I}A_{I}^{\pm} = \frac{1}{4\pi^2}$.
The resulting expression is
\begin{align}
\left(\langle \phi(\vec{x}) \phi(\vec{y}) \rangle^{\mathbb{RP}^d}\right)^{(1)}\epsilon = |\vec{x}-\vec{y}|^{-2} C_{\phi \phi}^{\ \ I}A_{I}^{\pm} \frac{(\gamma_{\phi^2})^{(1)}}{2}\epsilon \left[ \pm \frac{\eta}{1-\eta} \ln \eta + \ln (1-\eta) \right], \label{eq:2pt func in phi4 theory with perturbarion theory order epsilon}
\end{align}
which we will reproduce later from the other method.

\section{Crosscap bootstrap in order $\epsilon$} \label{Crosscap bootstrap in order epsilon}
As the second approach, in this section, we will solve the crosscap bootstrap equation\footnote{In \cite{Hogervorst:2017kbj}\cite{Hogervorst:2017sfd}, they investigated a solution for relevant crossing kernels in the crosscap bootstrap equations in the so-called ``alpha space''.} analytically in the $4-\epsilon$ dimensional critical $\phi^4$ theory to the first non-trivial order in $\epsilon$.
As we will demonstrate in the following, we can solve the crosscap bootstrap equation for the two-point function of the lowest dimensional scalar operator $\phi$ with itself to the first non-trivial order in $\epsilon$ by summing up a finite number of scalar primary operators in the conformal partial wave decomposition. A priori, this is quite non-trivial, but we can attribute it to the two salient features of the critical $\phi^4$ theory. The first feature is that the anomalous dimension of $\phi$ starts from $O(\epsilon^2)$ rather than $O(\epsilon)$.
 The second feature is that the scalar OPE of $\phi$ with itself can be truncated to a finite sum: $[\phi] \times [\phi] = I + [\phi^2] + [\phi^4] + O(\epsilon^2)$.

We have a small comment on the second feature. Naively we expect that the scalar OPE of $[\phi] \times [\phi]$ contains all the towers of operators like ``$\Box^k \phi^2$" and ``$\Box^k \phi^4$"  even at $O(\epsilon)$.\footnote{These are schematic notations taken from \cite{Liendo:2012hy}. What we really mean is that ``$\Box^k \phi^2$" and ``$\Box^k \phi^4$" are Lorentz scalar operators with two or four $\phi$s and $2k$ and $4k$ $\partial$s such as $\partial^\mu \phi \partial_\mu \phi$ for ``$\Box^{k=1} \phi^2$" and $\phi^2\partial_\mu \phi \partial^\mu \phi$ for ``$\Box^{k=1}\phi^4$". Note that at $k=1$, they are all descendant.}
 However, it turned out that the former behaves as conformal descendant operators, and we can ignore. The latter contains primary operators but they have remarkable properties that the OPE coefficient $C_{\phi\phi``\Box^{k=1}\phi^4"}$  is always $O(\epsilon^2)$. This was first noticed in solving the boundary bootstrap program in \cite{Liendo:2012hy} and further discussed from the large spin expansions in \cite{Alday:2017zzv}. We present the explicit computation of the vanishing OPE coefficient at $k=2$ in appendix \ref{phi-phi-Box2 phi4 three point function}.

Let us consider the crosscap bootstrap equation for the two-point functions of the lowest dimensional scalar operator $\phi$ with the conformal dimension $\Delta_{\phi}$ in $d=4-\epsilon$ dimensions:
\begin{align}
G_{\phi \phi}^{\pm}(\eta) = \pm \left( \frac{\eta}{1-\eta} \right)^{\Delta_{\phi}} G_{\phi \phi}^{\pm}(1-\eta), \label{eq:crosscap bootstrap equation in phi4 theory}
\end{align}
where
\begin{align}
\langle \phi(\vec{x}) \phi(\vec{y}) \rangle^{\mathbb{RP}^d} = |\vec{x}-\vec{y}|^{-2 \Delta_{\phi}} G_{\phi \phi}^{\pm}(\eta). \label{eq:phi phi 2pt func in phi4 theory}
\end{align}
The sign $\pm$ corresponds to the choice of the involution $\Omega$.

To implement and solve the crosscap bootstrap equation explicitly, we use the conformal partial wave decomposition
\begin{align}
G_{\phi \phi}^{\pm}(\eta) &= \sum_{\mathcal{O}=I,\phi^2,\phi^4,\cdots} C_{\phi \phi}^{\ \ \mathcal{O}}A_{\mathcal{O}}^{\pm} \eta^{\frac{\Delta_{\mathcal{O}}}{2}} {}_2F_1\left( \frac{\Delta_{\mathcal{O}}}{2}, \frac{\Delta_{\mathcal{O}}}{2}; \Delta_{\mathcal{O}}+1-\frac{d}{2}; \eta \right). \label{eq:CPWD in pht4 theory}
\end{align}
As we have already mentioned, we can truncate the sum only over three scalar primary operators  (i.e. $I$, $\phi^2$, and $\phi^4$).

To go further, we expand all the CFT data in power series of $\epsilon$. 
For the conformal dimension of scalar primary operator $\phi$, $\phi^2$, and $\phi^4$, we have
\begin{align}
&\Delta_{\phi} = \frac{d-2}{2} + \gamma_{\phi} = 1 - \frac{\epsilon}{2} + (\gamma_{\phi})^{(1)}\epsilon + O(\epsilon^2), \label{eq:epsilon expanded delta phi} \\
&\Delta_{\phi^2} = 2 - \epsilon + (\gamma_{\phi^2})^{(1)}\epsilon + O(\epsilon^2), \label{eq:epsilon expanded delta phi2} \\
&\Delta_{\phi^4} = 4 - 2 \epsilon + (\gamma_{\phi^4})^{(1)}\epsilon + O(\epsilon^2), \label{eq:epsilon expanded delta phi4}
\end{align}
and for the products of the OPE coefficient and the one-point function coefficient, we have
\begin{align}
&C_{\phi \phi}^{\ \ I} A_{I}^{\pm} = \frac{1}{4\pi^2} : \mathrm{normalization}, \label{eq: C phi phi I A I} \\
&C_{\phi \phi}^{\ \ \phi^2} A_{\phi^2}^{\pm} = (C_{\phi \phi}^{\ \ \phi^2} A_{\phi^2}^{\pm})^{(0)} + (C_{\phi \phi}^{\ \ \phi^2} A_{\phi^2}^{\pm})^{(1)} \epsilon + O(\epsilon^2), \label{eq:epsilon expanded C phi phi phi2 A phi2} \\
&C_{\phi \phi}^{\ \ \phi^4} A_{\phi^4}^{\pm} = (C_{\phi \phi}^{\ \ \phi^4} A_{\phi^4}^{\pm})^{(1)} \epsilon + O(\epsilon^2). \label{eq:epsilon expanded C phi phi phi4 A phi4}
\end{align}

Let us now solve the crosscap bootstrap equation by substituting $\epsilon$-expanded CFT data into the crosscap bootstrap equation to the first non-trivial order in $\epsilon$.
Concretely, substituting \eqref{eq:epsilon expanded delta phi}, \eqref{eq:epsilon expanded delta phi2}, \eqref{eq:epsilon expanded delta phi4}, \eqref{eq:epsilon expanded C phi phi phi2 A phi2}, and \eqref{eq:epsilon expanded C phi phi phi4 A phi4} into  \eqref{eq:crosscap bootstrap equation in phi4 theory} we obtain
\begin{align}
&C_{\phi \phi}^{\ \ I}A_{I}^{\pm} + C_{\phi \phi}^{\ \ \phi^2}A_{\phi^2}^{\pm} \eta^{\frac{\Delta_{\phi^2}}{2}} {}_2F_1\left( \frac{\Delta_{\phi^2}}{2}, \frac{\Delta_{\phi^2}}{2}; \Delta_{\phi^2}+1-\frac{d}{2}; \eta \right) \nonumber \\
&\quad + C_{\phi \phi}^{\ \ \phi^4}A_{\phi^4}^{\pm} \eta^{\frac{\Delta_{\phi^4}}{2}} {}_2F_1\left( \frac{\Delta_{\phi^4}}{2}, \frac{\Delta_{\phi^4}}{2}; \Delta_{\phi^4}+1-\frac{d}{2}; \eta \right) \nonumber \\
&= \pm \left( \frac{\eta}{1-\eta} \right)^{\Delta_{\phi}} \left[ C_{\phi \phi}^{\ \ I}A_{I}^{\pm} + C_{\phi \phi}^{\ \ \phi^2}A_{\phi^2}^{\pm} (1-\eta)^{\frac{\Delta_{\phi^2}}{2}} {}_2F_1\left( \frac{\Delta_{\phi^2}}{2}, \frac{\Delta_{\phi^2}}{2}; \Delta_{\phi^2}+1-\frac{d}{2}; 1-\eta \right) \right. \nonumber \\
&\left. \quad + C_{\phi \phi}^{\ \ \phi^4}A_{\phi^4}^{\pm} (1-\eta)^{\frac{\Delta_{\phi^4}}{2}} {}_2F_1\left( \frac{\Delta_{\phi^4}}{2}, \frac{\Delta_{\phi^4}}{2}; \Delta_{\phi^4}+1-\frac{d}{2}; 1-\eta \right) \right], \label{eq:crosscap bootstrap equation in phi4 theory 2}
\end{align}
and by using formulae of Gauss's hypergeometric function (see appendix \ref{Formulae for Gauss's hypergeometric function}), we obtain
\begin{align}
(\mathrm{LHS}\eqref{eq:crosscap bootstrap equation in phi4 theory 2}) &= C_{\phi \phi}^{\ \ I} A_{I}^{\pm} \nonumber \\
& \quad + \left[(C_{\phi \phi}^{\ \ \phi^2} A_{\phi^2}^{\pm})^{(0)} + (C_{\phi \phi}^{\ \ \phi^2} A_{\phi^2}^{\pm})^{(1)} \epsilon \right] \eta^{1-\frac{\epsilon}{2} + \frac{(\gamma_{\phi^2})^{(1)}}{2}\epsilon} \nonumber \\
& \qquad \times {}_2 F_1 \left( 1 - \frac{\epsilon}{2} + \frac{(\gamma_{\phi^2})^{(1)}}{2}\epsilon,1-\frac{\epsilon}{2} + \frac{(\gamma_{\phi^2})^{(1)}}{2}\epsilon;1-\frac{\epsilon}{2} + (\gamma_{\phi^2})^{(1)}\epsilon;\eta \right) \nonumber \\
& \quad + (C_{\phi \phi}^{\ \ \phi^4} A_{\phi^4}^{\pm})^{(1)} \epsilon \cdot \eta^{2}{}_2 F_1 \left(2,2;3;\eta \right) + O(\epsilon^2) \nonumber \\
&= C_{\phi \phi}^{\ \ I} A_{I}^{\pm} + (C_{\phi \phi}^{\ \ \phi^2} A_{\phi^2}^{\pm})^{(0)} \left( \frac{\eta}{1-\eta} \right)^{1-\frac{\epsilon}{2}} \nonumber \\
& \quad + (C_{\phi \phi}^{\ \ \phi^2} A_{\phi^2}^{\pm})^{(0)} \frac{(\gamma_{\phi^2})^{(1)}}{2}\epsilon \frac{\eta}{1-\eta} \ln \eta + (C_{\phi \phi}^{\ \ \phi^2} A_{\phi^2}^{\pm})^{(1)} \epsilon \frac{\eta}{1-\eta} \nonumber \\
& \quad + 2 (C_{\phi \phi}^{\ \ \phi^4} A_{\phi^4}^{\pm})^{(1)} \epsilon \ln(1-\eta) + 2 (C_{\phi \phi}^{\ \ \phi^4} A_{\phi^4}^{\pm})^{(1)} \epsilon \frac{\eta}{1-\eta} + O(\epsilon^2),
\end{align}
and
\begin{align}
(\mathrm{RHS}\eqref{eq:crosscap bootstrap equation in phi4 theory 2}) &= \pm \left( \frac{\eta}{1-\eta}\right)^{1-\frac{\epsilon}{2} + (\gamma_{\phi})^{(1)}\epsilon} \left[ C_{\phi \phi}^{\ \ I} A_{I}^{\pm} \right. \nonumber \\
& \quad + \left[(C_{\phi \phi}^{\ \ \phi^2} A_{\phi^2}^{\pm})^{(0)} + (C_{\phi \phi}^{\ \ \phi^2} A_{\phi^2}^{\pm})^{(1)} \epsilon \right] (1-\eta)^{1-\frac{\epsilon}{2} + \frac{(\gamma_{\phi^2})^{(1)}}{2}\epsilon} \nonumber \\
& \qquad \times {}_2 F_1 \left( 1 - \frac{\epsilon}{2} + \frac{(\gamma_{\phi^2})^{(1)}}{2}\epsilon,1-\frac{\epsilon}{2} + \frac{(\gamma_{\phi^2})^{(1)}}{2}\epsilon;1-\frac{\epsilon}{2} + (\gamma_{\phi^2})^{(1)}\epsilon;1-\eta \right) \nonumber \\
& \left. \quad + (C_{\phi \phi}^{\ \ \phi^4} A_{\phi^4}^{\pm})^{(1)} \epsilon (1-\eta)^{2}{}_2 F_1 \left(2,2;3;1-\eta \right) \right]  \nonumber \\
&= \pm C_{\phi \phi}^{\ \ I} A_{I}^{\pm} \left( \frac{\eta}{1-\eta}\right)^{1-\frac{\epsilon}{2} + (\gamma_{\phi})^{(1)}\epsilon} \pm (C_{\phi \phi}^{\ \ \phi^2} A_{\phi^2}^{\pm})^{(0)} \left( \frac{\eta}{1-\eta}\right)^{(\gamma_{\phi})^{(1)}\epsilon}  \nonumber \\
&\quad \pm (C_{\phi \phi}^{\ \ \phi^2} A_{\phi^2}^{\pm})^{(0)} \frac{(\gamma_{\phi^2})^{(1)}}{2}\epsilon \ln (1-\eta) \pm (C_{\phi \phi}^{\ \ \phi^2} A_{\phi^2}^{\pm})^{(1)} \epsilon \nonumber \\
&\quad \pm 2 (C_{\phi \phi}^{\ \ \phi^4} A_{\phi^4}^{\pm})^{(1)} \epsilon \frac{\eta}{1-\eta} \ln\eta\pm 2 (C_{\phi \phi}^{\ \ \phi^4} A_{\phi^4}^{\pm})^{(1)} \epsilon + O(\epsilon^2).
\end{align}

In the free theory limit (i.e. $\epsilon \to 0$), the crosscap bootstrap equation \eqref{eq:crosscap bootstrap equation in phi4 theory 2} becomes
\begin{align}
C_{\phi \phi}^{\ \ I}A_{I}^{\pm} + (C_{\phi \phi}^{\ \ \phi^2} A_{\phi^2}^{\pm})^{(0)} \frac{\eta}{1-\eta} = \pm C_{\phi \phi}^{\ \ I}A_{I}^{\pm} \frac{\eta}{1-\eta} \pm (C_{\phi \phi}^{\ \ \phi^2} A_{\phi^2}^{\pm})^{(0)}. \label{eq:cross cap bootstrap equation in free theory limit}
\end{align}
Comparing both sides of \eqref{eq:cross cap bootstrap equation in free theory limit}, we find
\begin{align}
(C_{\phi \phi}^{\ \ \phi^2} A_{\phi^2}^{\pm})^{(0)} = \pm C_{\phi \phi}^{\ \ I} A_{I}^{\pm} \label{eq:relationship of epsilon expnanded CFT data to O(1)}.
\end{align}
Let us substitute \eqref{eq:relationship of epsilon expnanded CFT data to O(1)}  back into  \eqref{eq:cross cap bootstrap equation in free theory limit} to rewrite the left-hand side of \eqref{eq:crosscap bootstrap equation in phi4 theory 2} and the right-hand side of \eqref{eq:crosscap bootstrap equation in phi4 theory 2} as
\begin{align}
(\mathrm{LHS}\eqref{eq:crosscap bootstrap equation in phi4 theory 2}) &= C_{\phi \phi}^{\ \ I} A_{I}^{\pm} \pm C_{\phi \phi}^{\ \ I} A_{I}^{\pm} \left( \frac{\eta}{1-\eta} \right)^{1-\frac{\epsilon}{2}} \nonumber \\
& \quad \pm C_{\phi \phi}^{\ \ I} A_{I}^{\pm} \frac{(\gamma_{\phi^2})^{(1)}}{2}\epsilon \frac{\eta}{1-\eta} \ln \eta + (C_{\phi \phi}^{\ \ \phi^2} A_{\phi^2}^{\pm})^{(1)} \epsilon \frac{\eta}{1-\eta} \nonumber \\
& \quad + 2 (C_{\phi \phi}^{\ \ \phi^4} A_{\phi^4}^{\pm})^{(1)} \epsilon \ln(1-\eta) + 2 (C_{\phi \phi}^{\ \ \phi^4} A_{\phi^4}^{\pm})^{(1)} \epsilon \frac{\eta}{1-\eta} + O(\epsilon^2) \nonumber \\
&= C_{\phi \phi}^{\ \ I} A_{I}^{\pm} \pm C_{\phi \phi}^{\ \ I} A_{I}^{\pm} \left( \frac{\eta}{1-\eta} \right) \pm \left( -\frac{\epsilon}{2} \right) C_{\phi \phi}^{\ \ I} A_{I}^{\pm} \ln \frac{\eta}{1-\eta} \nonumber \\
& \quad \pm C_{\phi \phi}^{\ \ I} A_{I}^{\pm} \frac{(\gamma_{\phi^2})^{(1)}}{2}\epsilon \frac{\eta}{1-\eta} \ln \eta + (C_{\phi \phi}^{\ \ \phi^2} A_{\phi^2}^{\pm})^{(1)} \epsilon \frac{\eta}{1-\eta} \nonumber \\
& \quad + 2 (C_{\phi \phi}^{\ \ \phi^4} A_{\phi^4}^{\pm})^{(1)} \epsilon \ln(1-\eta) + 2 (C_{\phi \phi}^{\ \ \phi^4} A_{\phi^4}^{\pm})^{(1)} \epsilon \frac{\eta}{1-\eta} + O(\epsilon^2) \ ,
\end{align}
and
\begin{align}
(\mathrm{RHS}\eqref{eq:crosscap bootstrap equation in phi4 theory 2}) &= \pm C_{\phi \phi}^{\ \ I} A_{I}^{\pm} \left( \frac{\eta}{1-\eta}\right)^{1-\frac{\epsilon}{2} + (\gamma_{\phi})^{(1)}\epsilon} + C_{\phi \phi}^{\ \ I} A_{I}^{\pm} \left( \frac{\eta}{1-\eta}\right)^{(\gamma_{\phi})^{(1)}\epsilon}  \nonumber \\
&\quad + C_{\phi \phi}^{\ \ I} A_{I}^{\pm} \frac{(\gamma_{\phi^2})^{(1)}}{2}\epsilon \ln (1-\eta) \pm (C_{\phi \phi}^{\ \ \phi^2} A_{\phi^2}^{\pm})^{(1)} \epsilon \nonumber \\
&\quad \pm 2 (C_{\phi \phi}^{\ \ \phi^4} A_{\phi^4}^{\pm})^{(1)} \epsilon \frac{\eta}{1-\eta} \ln\eta \pm 2 (C_{\phi \phi}^{\ \ \phi^4} A_{\phi^4}^{\pm})^{(1)} \epsilon + O(\epsilon^2) \nonumber \\
&= \pm C_{\phi \phi}^{\ \ I} A_{I}^{\pm} \left( \frac{\eta}{1-\eta}\right) \pm \left( -\frac{\epsilon}{2} + (\gamma_{\phi})^{(1)}\epsilon \right) C_{\phi \phi}^{\ \ I} A_{I}^{\pm} \ln \frac{\eta}{1-\eta} \nonumber \\
&\quad + C_{\phi \phi}^{\ \ I} A_{I}^{\pm} \left( \frac{\eta}{1-\eta}\right) + (\gamma_{\phi})^{(1)}\epsilon C_{\phi \phi}^{\ \ I} A_{I}^{\pm} \ln \frac{\eta}{1-\eta} \nonumber \\
&\quad + C_{\phi \phi}^{\ \ I} A_{I}^{\pm} \frac{(\gamma_{\phi^2})^{(1)}}{2}\epsilon \ln (1-\eta) \pm (C_{\phi \phi}^{\ \ \phi^2} A_{\phi^2}^{\pm})^{(1)} \epsilon \nonumber \\
&\quad \pm 2 (C_{\phi \phi}^{\ \ \phi^4} A_{\phi^4}^{\pm})^{(1)} \epsilon \frac{\eta}{1-\eta} \ln\eta \pm 2 (C_{\phi \phi}^{\ \ \phi^4} A_{\phi^4}^{\pm})^{(1)} \epsilon + O(\epsilon^2).
\end{align}

We now compare the coefficients of the terms that are the same functional form of $\eta$ on both sides at order $\epsilon$:
\begin{align}
&\epsilon \ln \frac{\eta}{1-\eta}: \pm \left( - \frac{\epsilon}{2} \right) = \pm \left( -\frac{\epsilon}{2} + (\gamma_{\phi})^{(1)}\epsilon \right), \\
&\epsilon \frac{\eta}{1-\eta} \ln \eta: \pm C_{\phi \phi}^{\ \ I} A_{I}^{\pm} \frac{(\gamma_{\phi^2})^{(1)}}{2}\epsilon = \pm 2 (C_{\phi \phi}^{\ \ \phi^4} A_{\phi^4}^{\pm})^{(1)} \epsilon,\\
&\epsilon \ln(1-\eta): 2 (C_{\phi \phi}^{\ \ \phi^4} A_{\phi^4}^{\pm})^{(1)} \epsilon = C_{\phi \phi}^{\ \ I} A_{I}^{\pm} \frac{(\gamma_{\phi^2})^{(1)}}{2}\epsilon, \\
&\epsilon \frac{\eta}{1-\eta}: (C_{\phi \phi}^{\ \ \phi^2} A_{\phi^2}^{\pm})^{(1)} \epsilon + 2 (C_{\phi \phi}^{\ \ \phi^4} A_{\phi^4}^{\pm})^{(1)} \epsilon=0, \\
&\epsilon \eta^0: 0 = \pm (C_{\phi \phi}^{\ \ \phi^2} A_{\phi^2}^{\pm})^{(1)} \epsilon \pm 2 (C_{\phi \phi}^{\ \ \phi^4} A_{\phi^4}^{\pm})^{(1)} \epsilon,
\end{align}
which gives the following necessary conditions among the CFT data:
\begin{align}
&(\gamma_{\phi})^{(1)} = 0, \label{eq:relationship of epsilon expnanded CFT data to O(epsilon) gamma phi} \\
&(\gamma_{\phi^2})^{(1)} = \frac{4 (C_{\phi \phi}^{\ \ \phi^4} A_{\phi^4}^{\pm})^{(1)}}{C_{\phi \phi}^{\ \ I} A_{I}^{\pm}}, \label{eq:relationship of epsilon expnanded CFT data to O(epsilon) gamma phi2} \\
&(C_{\phi \phi}^{\ \ \phi^2} A_{\phi^2}^{\pm})^{(1)} = - 2 (C_{\phi \phi}^{\ \ \phi^4} A_{\phi^4}^{\pm})^{(1)}. \label{eq:relationship of epsilon expnanded CFT data to O(epsilon) C phi phi phi2 A phi2}
\end{align}
Note that the relation \eqref{eq:relationship of epsilon expnanded CFT data to O(epsilon) gamma phi} is consistent with the fact that the anomalous dimension $\gamma_{\phi}$ starts from order $\epsilon^2$ in perturbation theory (i.e. $\gamma_{\phi}= \frac{1}{108}\epsilon^2 + O(\epsilon^3)$).

Finally, by substituting these conditions into \eqref{eq:crosscap bootstrap equation in phi4 theory 2}, we obtain the crossing symmetric solution of the crosscap bootstrap equation at $O(\epsilon)$:
\begin{align}
G_{\phi \phi}^{\pm}(\eta) = C_{\phi \phi}^{\ \ I} A_{I}^{\pm} \left[ 1 \pm \left( \frac{\eta}{1-\eta} \right)^{1-\frac{\epsilon}{2}} + \frac{(\gamma_{\phi^2})^{(1)}}{2}\epsilon \left[ \pm \frac{\eta}{1-\eta} \ln \eta + \ln (1-\eta)\right] \right] + O(\epsilon^2). \label{eq:order epsilon G phi phi pm}
\end{align}
This agrees with the perturbative computation \eqref{eq:2pt func in phi4 theory with perturbarion theory order epsilon} in the previous section.

We have a couple of comments here. First of all, the existence of the crossing symmetric solution a posteriori justifies our truncated scalar OPE ansatz  $[\phi] \times [\phi] = I + [\phi^2] + [\phi^4] + O(\epsilon^2)$. Secondly, the solution contains one free parameter $(\gamma_{\phi^2})^{(1)}$, which remains undetermined from the bootstrap approach taken here. There is a simple explanation of this. Consider the $O(N)$ critical vector models instead of the critical $\phi^4$ theory and study the crosscap bootstrap equation for two-point functions of scalar operators in the $O(N)$ vector representation i.e. $\langle \phi^I(x) \phi^J(y) \rangle$. What we obtain is the same crosscap bootstrap equation as in the critical $\phi^4$ theory, so the solution of the our crosscap bootstrap equation should contain one free parameter corresponding to $N$ e.g. appearing in $(\gamma_{\phi^2})^{(1)} = \frac{N+2}{N+8}$.

\section{Schwinger-Dyson equation approach}\label{Schwinger-Dyson equation approach}

As the third approach, in this section, we derive the CFT data that appears in the $\phi$-$\phi$ two-point function in the critical $\phi^4$ theory by using the conformal symmetry and the Schwinger-Dyson equations. The method proposed in \cite{Rychkov:2015naa} was to compute the CFT data in $\epsilon$ expansions without using the Feynman diagrams, but using the structure of the recombination of conformal multiplets (see also \cite{Sen:2015doa}). Later in \cite{Nii:2016lpa}, the more active use of the equations of motion is advocated. The spirit of our approach is closer to the latter.

Let us first recall the equations of motion for the elementary scalar field in the critical $\phi^4$ theory. Inside any correlation functions, we expect that the  equation of motion obtained from varying $\phi$ in the classical action holds:
\begin{align}
\langle \Box_{x} \phi (\vec{x})  \cdots \rangle^{\mathbb{RP}^d} = \langle \frac{g}{3!} \phi^3 (\vec{x}) \cdots \rangle^{\mathbb{RP}^d}, \label{eq:EoM of phi in ctirical phi4 theory}
\end{align}
where $\Box:=\partial^2$ is $d$-dimensional Laplacian. Such equations of motion should hold not only on ${\mathbb{R}^d}$ but also on ${\mathbb{RP}^d}$. This  is a concrete realization of the multiplet recombination phenomenon in \cite{Rychkov:2015naa}.\footnote{See also \cite{Gliozzi:2016ysv}\cite{Gliozzi:2017hni}\cite{Gliozzi:2017gzh}\cite{Safari:2017tgs} for  the related studies of multiplet recombinations on ${\mathbb{R}^d}$.}

More formally, by using the change of variable $\phi \to \phi + \delta \phi$ in the path integral expression for the one-point function of $\langle \phi^n(\vec{x}) \rangle^{\mathbb{RP}^d}$ on $\mathbb{RP}^d$, we obtain
\begin{align}
\langle n\phi^{n-1}(\vec{x}) \delta \phi(\vec{x}) \rangle^{\mathbb{RP}^d} = \langle \phi^n(\vec{x}) \int \mathrm{d}^d z \delta\phi(\vec{z}) (\Box_{z} \phi (\vec{z})-  \frac{g}{3!} \phi^3 (\vec{z})) \rangle^{\mathbb{RP}^d} \ .
\end{align}
If we set $\delta \phi(\vec{z}) = \delta(\vec{z}-\vec{y})$, this becomes 
\begin{align}
\langle \phi^n(\vec{x}) (\Box_{y} \phi (\vec{y})-  \frac{g}{3!} \phi^3 (\vec{y})) \rangle^{\mathbb{RP}^d} = \delta(\vec{x}-\vec{y}) \langle n\phi^{n-1}(\vec{x}) \rangle^{\mathbb{RP}^d}  \ . \label{SD1}
\end{align}
We will neglect the contact terms appearing in the right-hand side. Similarly, starting with the one-point function $\langle \Box_x\phi(\vec{x}) \rangle^{\mathbb{RP}^d}$, we obtain
\begin{align}
\langle \Box_x \phi(\vec{x}) \Box_{y} \phi (\vec{y}) \rangle^{\mathbb{RP}^d} =  \langle \Box_x \phi(\vec{x}) \frac{g}{3!} \phi^3 (\vec{y}) \rangle^{\mathbb{RP}^d} =  \langle \frac{g}{3!} \phi^3 (\vec{x})  \frac{g}{3!} \phi^3 (\vec{y}) \rangle^{\mathbb{RP}^d}  \label{SD2}
\end{align}
up to a contact term. We are going to use \eqref{SD1} and \eqref{SD2} to solve the CFT data in the critical $\phi^4$ theory in the following.

This perturbative picture allows us to write down three axioms in the $\epsilon$-expansion from CFT \cite{Rychkov:2015naa} to define (and solve) the critical $\phi^4$ theory with the Wilson-Fisher fixed point:
\begin{description}
\item[Axiom I] \mbox{} 
The Wilson-Fisher fixed point has conformal symmetry.
\item[Axiom II] \mbox{} 
If we take the $\epsilon \rightarrow 0$ limit, correlation functions in the interacting theory will approach the ones in the free theory.
\item[Axiom III] \mbox{} 
From the Schwinger-Dyson equation, a particular primary operator in the free theory (i.e. $\phi^{3}$) behaves as a descendant operator at the Wilson-Fisher fixed point (i.e. $\phi^{3}$ is the descendant of $\phi$ by acting the Laplacian as in \eqref{eq:EoM of phi in ctirical phi4 theory}).
\end{description}

Let us begin with the normalization of the two-point function. We recall that we have fixed the normalization of the two-point function  at the Gaussian fixed point as in \eqref{eq:1pt func in 4-epsilon g=0 phi4 theory}, \eqref{eq:phi2 correlation func in 4-epsilon g=0 phi4 theory}, and \eqref{eq:2pt func in 4-epsilon g=0 phi4 theory}. The simplest way to use the condition is to take the free theory limit of the $\phi$-$\phi$ two-point function in the critical $\phi^4$ theory as
\begin{align}
\lim_{\epsilon \to 0} \langle \phi (\vec{x}) \phi (\vec{y}) \rangle^{\mathbb{RP}^d} = \langle \phi (\vec{x}) \phi (\vec{y}) \rangle^{\mathbb{RP}^d}_{\mathrm{free}}, \label{eq:free theory limit for 2pt function}
\end{align}
where we can evaluate the left-hand side by using the conformal partial wave decomposition:
\begin{align}
(\mathrm{LHS} \eqref{eq:free theory limit for 2pt function}) &= |\vec{x}-\vec{y}|^{-2} \left[ C_{\phi \phi}^{\ \ I}A_{I}^{\pm}  + (C_{\phi \phi}^{\ \ \phi^2}A_{\phi^2}^{\pm})^{(0)} \eta \cdot {}_2F_1\left( 1, 1; 1; \eta \right)  \right] \nonumber \\
&= |\vec{x}-\vec{y}|^{-2} \left[ C_{\phi \phi}^{\ \ I}A_{I}^{\pm}  + (C_{\phi \phi}^{\ \ \phi^2}A_{\phi^2}^{\pm})^{(0)} \left( \frac{\eta}{1-\eta} \right)  \right],
\end{align}
and we can evaluate the right-hand side in the free field theory:
\begin{align}
(\mathrm{RHS} \eqref{eq:free theory limit for 2pt function}) &= \frac{1}{4 \pi^2} \frac{1}{|\vec{x}-\vec{y}|^2} \left[ 1 \pm \left( \frac{\eta}{1-\eta} \right) \right].
\end{align}
Comparing the coefficients of the terms that are the same functional form of $\eta$ on both sides, we find
\begin{align}
&C_{\phi \phi}^{\ \ I} A_{I}^{\pm} = \frac{1}{4\pi^2} : \mathrm{normalization}, \label{eq:C phi phi I A I in critical phi4 theory} \\
&(C_{\phi \phi}^{\ \ \phi^2} A_{\phi^2}^{\pm})^{(0)}  = \pm \frac{1}{4\pi^2} (= \pm C_{\phi \phi}^{\ \ I} A_{I}^{\pm} ). \label{eq:C phi phi phi2 A phi2 in critical phi4 theory}
\end{align}

To obtain more non-trivial results, we are going to act the Laplacian on the $\phi$-$\phi$ two-point function. The form of the $\phi$-$\phi$ two-point function is fixed by the conformal invariance (axiom I), and we apply the Schwinger-Dyson equation \eqref{eq:EoM of phi in ctirical phi4 theory} as axiom III:
\begin{align}
\langle \Box_x \phi(\vec{x}) \phi(\vec{y}) \rangle^{\mathbb{RP}^d} = \frac{g}{3!} \langle \phi^3(\vec{x}) \phi(\vec{y}) \rangle^{\mathbb{RP}^d}. \label{eq: Laplacian acting once 2pt func phi phi in phi4 theory}
\end{align}
Then, we take the $\epsilon \to 0$ limit from axiom II to evaluate the right-hand side in the free field theory. This gives us non-trivial consistency conditions at the first order in $\epsilon$.

Explicitly, for the left-hand side of \eqref{eq: Laplacian acting once 2pt func phi phi in phi4 theory}, we know the concrete form of the two-point function, so we can just differentiate it (see appendix \ref{Laplacian acting two-point functions} for the computation)
\begin{align}
(\mathrm{LHS} \eqref{eq: Laplacian acting once 2pt func phi phi in phi4 theory}) &= \Box_{x} |\vec{x}-\vec{y}|^{-2 \Delta_{\phi}} \sum_{\mathcal{O}=I,\phi^2,\phi^4,\cdots} C_{\phi \phi}^{\ \ \mathcal{O}}A_{\mathcal{O}}^{\pm} \eta^{\frac{\Delta_{\mathcal{O}}}{2}} {}_2F_1\left( \frac{\Delta_{\mathcal{O}}}{2}, \frac{\Delta_{\mathcal{O}}}{2}; \Delta_{\mathcal{O}}+1-\frac{d}{2}; \eta \right) \nonumber \\
&= \Box_{x} |\vec{x}-\vec{y}|^{-2 \Delta_{\phi}} \left[ C_{\phi \phi}^{\ \ I} A_{I}^{\pm} \right. \nonumber \\
& \quad + \left[(C_{\phi \phi}^{\ \ \phi^2} A_{\phi^2}^{\pm})^{(0)} + (C_{\phi \phi}^{\ \ \phi^2} A_{\phi^2}^{\pm})^{(1)} \epsilon \right] \eta^{\frac{\Delta_{\phi^2}}{2}} \left( 1 + \left( 1 - \frac{\epsilon}{2} \right)\eta + O(\eta^2) \right) \nonumber \\
& \left. \quad + (C_{\phi \phi}^{\ \ \phi^4} A_{\phi^4}^{\pm})^{(1)} \epsilon \cdot \eta^{\frac{\Delta_{\phi^4}}{2}} (1 + O(\eta) ) + O(\epsilon^2) \right] \nonumber \\
&= \left[ 4 C_{\phi \phi}^{\ \ I} A_{I}^{\pm} (\gamma_{\phi})^{(1)}\epsilon |\vec{x}-\vec{y}|^{-4} \right. \nonumber \\
&\quad + 2 (C_{\phi \phi}^{\ \ \phi^2} A_{\phi^2}^{\pm})^{(0)} \left[ (\gamma_{\phi^2})^{(1)}\epsilon -2(\gamma_{\phi})^{(1)}\epsilon \right] \eta |\vec{x}-\vec{y}|^{-4} \nonumber \\
&\quad - 4 (C_{\phi \phi}^{\ \ \phi^2} A_{\phi^2}^{\pm})^{(0)} (\gamma_{\phi^2})^{(1)} \epsilon \frac{(\vec{x}-\vec{y}) \cdot \vec{x}}{1+|\vec{x}|^2} \eta |\vec{x}-\vec{y}|^{-4} \nonumber \\
&\quad - 8 (C_{\phi \phi}^{\ \ \phi^2} A_{\phi^2}^{\pm})^{(0)} \left[ 1 - \frac{3}{4}\epsilon + \frac{3}{4}(\gamma_{\phi^2})^{(1)}\epsilon \right] \eta^2 \frac{1+|\vec{y}|^2}{1+|\vec{x}|^2} |\vec{x}-\vec{y}|^{-4} \nonumber \\
&\quad + 2 (C_{\phi \phi}^{\ \ \phi^2} A_{\phi^2}^{\pm})^{(0)} (\gamma_{\phi^2})^{(1)}\epsilon (1+|\vec{y}|^2) \eta^2 |\vec{x}-\vec{y}|^{-4} \nonumber \\
&\quad + 8 (C_{\phi \phi}^{\ \ \phi^2} A_{\phi^2}^{\pm})^{(0)} \left[ 1 - \frac{3}{4}\epsilon + \frac{3}{4}(\gamma_{\phi^2})^{(1)}\epsilon -\frac{3}{2}(\gamma_{\phi})^{(1)}\epsilon \right] \eta^2 |\vec{x}-\vec{y}|^{-4} \nonumber \\
&\quad - 8 (C_{\phi \phi}^{\ \ \phi^2} A_{\phi^2}^{\pm})^{(1)} \epsilon \eta^2 \frac{1+|\vec{y}|^2}{1+|\vec{x}|^2} |\vec{x}-\vec{y}|^{-4} \nonumber \\
&\quad + 8 (C_{\phi \phi}^{\ \ \phi^2} A_{\phi^2}^{\pm})^{(1)} \epsilon \eta^2 |\vec{x}-\vec{y}|^{-4} \nonumber \\
&\quad \left. + 8 (C_{\phi \phi}^{\ \ \phi^4} A_{\phi^4}^{\pm})^{(1)} \epsilon \eta^2 |\vec{x}-\vec{y}|^{-4} \right] + O(\epsilon^2) \nonumber \\
&\sim \left[ 4 C_{\phi \phi}^{\ \ I} A_{I}^{\pm} (\gamma_{\phi})^{(1)}\epsilon |\vec{x}-\vec{y}|^{-4} \right. \nonumber \\
&\quad + 2 (C_{\phi \phi}^{\ \ \phi^2} A_{\phi^2}^{\pm})^{(0)} \left[ (\gamma_{\phi^2})^{(1)} -2(\gamma_{\phi})^{(1)} \right] \epsilon \eta |\vec{x}-\vec{y}|^{-4} \nonumber \\
&\quad - 2 (C_{\phi \phi}^{\ \ \phi^2} A_{\phi^2}^{\pm})^{(0)} (\gamma_{\phi^2})^{(1)} \epsilon (|\vec{x}|^2-|\vec{y}|^2) \eta |\vec{x}-\vec{y}|^{-4} \nonumber \\
&\quad - 12(C_{\phi \phi}^{\ \ \phi^2} A_{\phi^2}^{\pm})^{(0)} (\gamma_{\phi})^{(1)}\epsilon \eta^2 |\vec{x}-\vec{y}|^{-4} \nonumber \\
&\quad \left. + 8 (C_{\phi \phi}^{\ \ \phi^4} A_{\phi^4}^{\pm})^{(1)} \epsilon \eta^2 |\vec{x}-\vec{y}|^{-4} \right] + O(\epsilon^2) \ . \label{eq:LHS once Laplacian acting 2pt func for phi}
\end{align}
In the last line, we have expanded the results around $x=y=\eta=0$ to simplify our comparison in the following. For this purpose, we have used the identity $\frac{(\vec{x}-\vec{y}) \cdot \vec{x}}{1+|\vec{x}|^2} = \frac{1}{2} \eta (1 + |\vec{y}|^2) + \frac{1}{2} \frac{|\vec{x}|^2-|\vec{y}|^2}{1+|\vec{x}|^{2}}$.
For the right-hand side of \eqref{eq: Laplacian acting once 2pt func phi phi in phi4 theory}, since the prefactor $g = O (\epsilon)$ is multiplied, we can substitute the two-point function of the free-field theory:
\begin{align}
(\mathrm{RHS} \eqref{eq: Laplacian acting once 2pt func phi phi in phi4 theory}) &\sim \frac{g}{3!} \langle \phi^3(\vec{x}) \phi(\vec{y}) \rangle^{\mathbb{RP}^d}_{\mathrm{free}} \nonumber \\
&= \frac{g}{3!} \cdot 3 \langle \phi(\vec{x}) \phi(\vec{y}) \rangle^{\mathbb{RP}^d}_{\mathrm{free}} \langle \phi^2(\vec{x}) \rangle^{\mathbb{RP}^d}_{\mathrm{free}} \nonumber \\
&= \pm \frac{g}{2} \left( \frac{1}{4\pi^2} \right)^2 \frac{(1+|\vec{x}|^2)^{-1}(1+|\vec{y}|^2)}{|\vec{x}-\vec{y}|^4} \left[ \eta \pm \eta \left( \frac{\eta}{1-\eta} \right) \right] \nonumber \\
&\sim \pm \frac{g}{2} \left( \frac{1}{4\pi^2} \right)^2 |\vec{x}-\vec{y}|^{-4} \left[ (1-|\vec{x}|^2+|\vec{y}|^2)\eta \pm \eta^2 + O(\eta^3) \right].
\end{align}
Again as in the left-hand side, we have expanded it around $x=y= \eta= 0$ to simplify the comparison.

Now if we compare the both sides of \eqref{eq: Laplacian acting once 2pt func phi phi in phi4 theory} at order $\epsilon$, we obtain
\begin{align}
&O(\epsilon \eta^0): 4 (\gamma_{\phi})^{(1)} \epsilon |\vec{x}-\vec{y}|^{-4}=0, \\
&O(\epsilon \eta^1): 2 (C_{\phi \phi}^{\ \ \phi^2} A_{\phi^2}^{\pm})^{(0)} \left[ (1-|\vec{x}|^2+|\vec{y}|^2) (\gamma_{\phi^2})^{(1)}-2(\gamma_{\phi})^{(1)} \right] \epsilon |\vec{x}-\vec{y}|^{-4} \eta \nonumber \\
&\qquad \qquad = \pm \frac{g}{2} \left( \frac{1}{4\pi^2} \right)^2 (1-|\vec{x}|^2+|\vec{y}|^2) |\vec{x}-\vec{y}|^{-4} \eta, \\
&O(\epsilon \eta^2): \left[ - 12(C_{\phi \phi}^{\ \ \phi^2} A_{\phi^2}^{\pm})^{(0)} (\gamma_{\phi})^{(1)}\epsilon + 8 (C_{\phi \phi}^{\ \ \phi^4} A_{\phi^4}^{\pm})^{(1)} \epsilon \right] |\vec{x}-\vec{y}|^{-4} \eta^2 = \frac{g}{2} \left( \frac{1}{4\pi^2} \right)^2 |\vec{x}-\vec{y}|^{-4} \eta^2.
\end{align}
Solving these equations, we can determine the anomalous dimension of the lowest dimensional scalar $\phi$, the anomalous dimension of the next-lowest dimensional scalar $\phi^2$, and the quantity $C_{\phi \phi}^{\ \ \phi^4} A_{\phi^4}^{\pm}$ at $O(\epsilon)$:
\begin{align}
&(\gamma_{\phi})^{(1)}  = 0, \label{eq:anomalous dimension phi order epsilon 1} \\
&(\gamma_{\phi^2})^{(1)}\epsilon = \frac{g}{16\pi^2}, \label{eq:anomalous dimension phi2 in critical phi4 theory} \\
&(C_{\phi \phi}^{\ \ \phi^4} A_{\phi^4}^{\pm})^{(1)}\epsilon = \frac{g}{16} (C_{\phi \phi}^{\ \ I} A_{I}^{\pm})^2, \label{eq:C phi phi pih4 A phi4 in critical phi4 theory}
\end{align}
where $C_{\phi \phi}^{\ \ I} A_{I}^{\pm} = \frac{1}{4\pi^2}$.
Note that we have used \eqref{eq:C phi phi phi2 A phi2 in critical phi4 theory} and \eqref{eq:anomalous dimension phi order epsilon 1} to obtain \eqref{eq:anomalous dimension phi2 in critical phi4 theory}, and we have also used \eqref{eq:anomalous dimension phi order epsilon 1} to obtain \eqref{eq:C phi phi pih4 A phi4 in critical phi4 theory}.\footnote{
We could have simplified the above calculation by using the $SO(d+1)$ symmetry to move $\vec{y}$ to the origin. The result, completely in agreement with the one here, is reported in appendix \ref{Laplacian with y=0}.} In principle, there could have existed a contribution from $(C_{\phi \phi}^{\ \ \phi^2} A_{\phi^2}^{\pm})^{(1)}\epsilon$, but this did not appear, so one cannot determine the quantity $(C_{\phi \phi}^{\ \ \phi^2} A_{\phi^2}^{\pm})^{(1)}\epsilon$  in this approach.

One advantage of the Schwinger-Dyson approach is that we can easily study the next order in $\epsilon$. To do this we simply act the Laplacian twice on the $\phi$-$\phi$ two-point function. 
The Schwinger-Dyson equation  \eqref{eq:EoM of phi in ctirical phi4 theory} as axiom III becomes
\begin{align}
\langle \Box_x \phi(\vec{x}) \Box_y \phi(\vec{y}) \rangle^{\mathbb{RP}^d} = \frac{g^2}{(3!)^2} \langle \phi^3(\vec{x}) \phi^3(\vec{y}) \rangle^{\mathbb{RP}^d}. \label{eq: Laplacian acting twice 2pt func phi phi in phi4 theory}
\end{align}
For the left-hand side of \eqref{eq: Laplacian acting twice 2pt func phi phi in phi4 theory}, since we know the concrete form of the two-point function, we can just differentiate it:
\begin{align}
(\mathrm{LHS} \eqref{eq: Laplacian acting twice 2pt func phi phi in phi4 theory}) &= \Box_{x} \Box_{y} |\vec{x}-\vec{y}|^{-2 \Delta_{\phi}} \sum_{\mathcal{O}=I,\phi^2,\phi^4,\cdots} C_{\phi \phi}^{\ \ \mathcal{O}}A_{\mathcal{O}}^{\pm} \eta^{\frac{\Delta_{\mathcal{O}}}{2}} {}_2F_1\left( \frac{\Delta_{\mathcal{O}}}{2}, \frac{\Delta_{\mathcal{O}}}{2}; \Delta_{\mathcal{O}}+1-\frac{d}{2}; \eta \right) \\
&= \Box_{x} \Box_{y} |\vec{x}-\vec{y}|^{-2 \Delta_{\phi}} \left[ C_{\phi \phi}^{\ \ I} A_{I}^{\pm} \right. \nonumber \\
& \quad + \left[(C_{\phi \phi}^{\ \ \phi^2} A_{\phi^2}^{\pm})^{(0)} + (C_{\phi \phi}^{\ \ \phi^2} A_{\phi^2}^{\pm})^{(1)} \epsilon \right] \eta^{\frac{\Delta_{\phi^2}}{2}} \left( 1 + \left( 1 - \frac{\epsilon}{2} \right)\eta + O(\eta^2) \right) \nonumber \\
& \left. \quad + (C_{\phi \phi}^{\ \ \phi^4} A_{\phi^4}^{\pm})^{(1)} \epsilon \cdot \eta^{\frac{\Delta_{\phi^4}}{2}} (1 + O(\eta) ) + O(\epsilon^2) \right] \nonumber \\
&= C_{\phi \phi}^{\ \ I} A_{I}^{\pm} 2^5 \left[ (\gamma_{\phi})^{(1)}\epsilon + (\gamma_{\phi})^{(2)}\epsilon^2 \right] |\vec{x}-\vec{y}|^{-6} \nonumber \\
&\quad + 4 (C_{\phi \phi}^{\ \ \phi^2} A_{\phi^2}^{\pm})^{(0)} \left[ 2(\gamma_{\phi})^{(1)}\epsilon - (\gamma_{\phi^2})^{(1)}\epsilon \right] \left[ (\gamma_{\phi^2})^{(1)}\epsilon - 2(\gamma_{\phi})^{(1)}\epsilon - \epsilon \right] \eta |\vec{x}-\vec{y}|^{-6} \nonumber \\
&\quad + 16 (C_{\phi \phi}^{\ \ \phi^2} A_{\phi^2}^{\pm})^{(0)} \left[ 2(\gamma_{\phi})^{(1)}\epsilon - (\gamma_{\phi^2})^{(1)}\epsilon \right] \left[ 1 - \frac{3}{2}\epsilon + \frac{3}{2}(\gamma_{\phi^2})^{(1)}\epsilon -2 (\gamma_{\phi})^{(1)}\epsilon \right] \eta^2 |\vec{x}-\vec{y}|^{-6} \nonumber \\
&\quad + 16 (C_{\phi \phi}^{\ \ \phi^2} A_{\phi^2}^{\pm})^{(0)} \left[ (\gamma_{\phi^2})^{(1)}\epsilon - 2(\gamma_{\phi})^{(1)}\epsilon \right] \left[ 1 - \frac{5}{4}\epsilon + \frac{5}{4}(\gamma_{\phi^2})^{(1)}\epsilon - \frac{5}{2}(\gamma_{\phi})^{(1)}\epsilon \right] \eta^2 |\vec{x}-\vec{y}|^{-6}  \nonumber \\
&\quad + 16 (C_{\phi \phi}^{\ \ \phi^2} A_{\phi^2}^{\pm})^{(1)} \epsilon \left[ 2(\gamma_{\phi})^{(1)}\epsilon - (\gamma_{\phi^2})^{(1)}\epsilon \right] \eta^2 |\vec{x}-\vec{y}|^{-6} \nonumber \\
&\quad + 16 (C_{\phi \phi}^{\ \ \phi^2} A_{\phi^2}^{\pm})^{(1)} \epsilon \left[ (\gamma_{\phi^2})^{(1)}\epsilon - 2(\gamma_{\phi})^{(1)}\epsilon \right] \eta^2 |\vec{x}-\vec{y}|^{-6} \nonumber \\
&\quad + 16 (C_{\phi \phi}^{\ \ \phi^4} A_{\phi^4}^{\pm})^{(1)} \epsilon \left[ -\epsilon + (\gamma_{\phi^4})\epsilon -2(\gamma_{\phi})^{(1)}\epsilon \right] \eta^2 |\vec{x}-\vec{y}|^{-6} + O(\eta^3) + O(\epsilon^3) \nonumber \\
&= \left[ C_{\phi \phi}^{\ \ I} A_{I}^{\pm} 2^5 (\gamma_{\phi})^{(2)} + 4 (C_{\phi \phi}^{\ \ \phi^2} A_{\phi^2}^{\pm})^{(0)} (\gamma_{\phi^2})^{(1)}\left[ 1- (\gamma_{\phi^2})^{(1)} \right] \eta \right. \nonumber \\
&\quad \left. + \left[ 4 (C_{\phi \phi}^{\ \ \phi^2} A_{\phi^2}^{\pm})^{(0)} (\gamma_{\phi^2})^{(1)} \left[1- (\gamma_{\phi^2})^{(1)}  \right] + 16 (C_{\phi \phi}^{\ \ \phi^4} A_{\phi^4}^{\pm})^{(1)} [(\gamma_{\phi^4})^{(1)} - 1] \right] \eta^2  + O(\eta^3) \right] \nonumber \\
&\quad \times \epsilon^2 |\vec{x}-\vec{y}|^{-6} + O(\epsilon^3). \label{eq:LHS twice Laplacian acting 2pt func for phi}
\end{align}
In the last line, we have used the result of the anomalous dimension of the lowest dimensional scalar operator $\phi$ at order $\epsilon$ obtained in \eqref{eq:anomalous dimension phi order epsilon 1}.

For the right-hand side of \eqref{eq: Laplacian acting twice 2pt func phi phi in phi4 theory}, since the prefactor $g^2 \sim O (\epsilon^2)$ is multiplied, the two-point function on the Wilson-Fisher fixed point may be approximated by the correlation function of the free-field theory:
\begin{align}
(\mathrm{RHS} \eqref{eq: Laplacian acting twice 2pt func phi phi in phi4 theory}) &\sim \frac{g^2}{36} \langle \phi^3(\vec{x}) \phi^3(\vec{y}) \rangle^{\mathbb{RP}^d}_{\mathrm{free}} \nonumber \\
&= \frac{g^2}{36} \left[ 3! \cdot \left[ \langle \phi(\vec{x}) \phi(\vec{y}) \rangle^{\mathbb{RP}^d}_{\mathrm{free}} \right]^{3} + 9 \cdot \langle \phi(\vec{x}) \phi(\vec{y}) \rangle^{\mathbb{RP}^d}_{\mathrm{free}} \langle \phi^2(\vec{x}) \rangle^{\mathbb{RP}^d}_{\mathrm{free}}  \langle \phi^2(\vec{y}) \rangle^{\mathbb{RP}^d}_{\mathrm{free}} \right] \nonumber \\
&= \frac{g^2}{36} \left( \frac{1}{4\pi^2} \right)^3 \frac{1}{|\vec{x}-\vec{y}|^6} \left[ 6 \cdot \left[ 1 \pm \left( \frac{\eta}{1-\eta} \right) \right]^3 + 9 \cdot \left[ 1 \pm \left( \frac{\eta}{1-\eta} \right) \right] \eta^2 \right] \nonumber \\
&=
\begin{cases} 
\frac{g^2}{36} \left( \frac{1}{4\pi^2} \right)^3 \frac{1}{|\vec{x}-\vec{y}|^6} \left[ 6 + 18 \eta + 45 \eta^2 + O(\eta^3) \right] \nonumber \\
\frac{g^2}{36} \left( \frac{1}{4\pi^2} \right)^3 \frac{1}{|\vec{x}-\vec{y}|^6} \left[ 6 - 18 \eta + 9 \eta^2 + O(\eta^3) \right]. 
\end{cases}
\end{align}
Comparing both sides of \eqref{eq: Laplacian acting twice 2pt func phi phi in phi4 theory} at order $\epsilon^2$, we obtain
\begin{align}
&O(\epsilon^2 \eta^0): C_{\phi \phi}^{\ \ I} A_{I}^{\pm} 2^5 (\gamma_{\phi})^{(2)} \epsilon^2 |\vec{x}-\vec{y}|^{-6} = \frac{g^2}{6} \left( \frac{1}{4\pi^2} \right)^3 |\vec{x}-\vec{y}|^{-6}, \\
&O(\epsilon^2 \eta): 4 (C_{\phi \phi}^{\ \ \phi^2} A_{\phi^2}^{\pm})^{(0)} (\gamma_{\phi^2})^{(1)}\left[1- (\gamma_{\phi^2})^{(1)}\right]  \epsilon^2 |\vec{x}-\vec{y}|^{-6} \eta = \pm \frac{g^2}{2} \left( \frac{1}{4\pi^2} \right)^3 |\vec{x}-\vec{y}|^{-6} \eta, \\
&O(\epsilon^2 \eta^2): \left[ 4 (C_{\phi \phi}^{\ \ \phi^2} A_{\phi^2}^{\pm})^{(0)} (\gamma_{\phi^2})^{(1)}\left[1-(\gamma_{\phi^2})^{(1)}\right] + 16 (C_{\phi \phi}^{\ \ \phi^4} A_{\phi^4}^{\pm})^{(1)} \left[(\gamma_{\phi^4})^{(1)} - 1\right] \right] \epsilon^2 |\vec{x}-\vec{y}|^{-6} \eta^2 \nonumber \\
&\qquad \qquad =
\begin{cases}
\frac{5}{4} g^2 \left( \frac{1}{4\pi^2} \right)^3 |\vec{x}-\vec{y}|^{-6} \eta^2 \nonumber \\
\frac{1}{4} g^2 \left( \frac{1}{4\pi^2} \right)^3 |\vec{x}-\vec{y}|^{-6} \eta^2.
\end{cases}
\end{align}
Combining them with the previous order $\epsilon$ results, we can determine the anomalous dimension of the lowest dimensional scalar $\gamma_{\phi}$, the critical coupling $g_{*}$ , and the anomalous dimension of the third lowest dimensional scalar $\gamma_{\phi^4}$:
\begin{align}
&(\gamma_{\phi})^{(2)}\epsilon^{2} = \frac{g^2}{3 \cdot 4^3 \cdot (4\pi^2)^2}, \label{eq:anomalous dimension phi order epsilon 2} \\
&g_{*}=\frac{16 \pi^2}{3} \epsilon + O(\epsilon^2), \label{eq:critical coupling in critical phi4 theory} \\
&(\gamma_{\phi^4})^{(1)} \epsilon = 
\begin{cases}
6 \cdot \frac{g}{16 \pi^2} \label{eq:anomalous dimension phi 4} \\
2\epsilon.
\end{cases}
\end{align}
In particular, \eqref{eq:critical coupling in critical phi4 theory} specifies the critical coupling constant as a function of $\epsilon$. This is because we have demanded the conformal symmetry.

Summarizing the results for the CFT date in terms of $\epsilon$ without referring to the coupling constant $g$, we have
\begin{align}
&\gamma_{\phi} = \frac{1}{108}\epsilon^2 + O(\epsilon^3), \label{eq:gamma phi in critical phi4 theory in epsilon}  \\
&\gamma_{\phi^2} = \frac{1}{3}\epsilon + O(\epsilon^2), \label{eq:gamma phi 2 in critical phi4 theory in epsilon} \\
&\gamma_{\phi^4} = 2 \epsilon + O(\epsilon^2), \label{eq:gamma phi 4 in critical phi4 theory in epsilon} \\
&C_{\phi \phi}^{\ \ \phi^4} A_{\phi^4}^{\pm} = \frac{1}{4 \cdot 3 \cdot 4\pi^2}\epsilon + O(\epsilon^2). \label{eq:C phi phi phi4 A phi4 in critical phi4 theory in epsilon}
\end{align}
We emphasize again that although in principle we could have obtained $(C_{\phi \phi}^{\ \ \phi^2}A_{\phi^2}^{\pm})^{(1)}\epsilon$, it did not appear in the Schwinger-Dyson equation, so its value is not fixed in this approach.
The results of the anomalous dimension \eqref{eq:gamma phi in critical phi4 theory in epsilon}, \eqref{eq:gamma phi 2 in critical phi4 theory in epsilon}, and \eqref{eq:gamma phi 4 in critical phi4 theory in epsilon} are in agreement with the known results in perturbation theory computed in $\mathbb{R}^d$ \cite{Rychkov:2015naa}\cite{Nii:2016lpa}. In addition, we have determined the additional CFT data $C_{\phi \phi}^{\ \ \phi^4} A_{\phi^4}^{\pm}$  on $\mathbb{RP}^d$ from the Schwinger-Dyson equation with the conformal symmetry.

\section{Conclusion} \label{Conclusion}
In this paper, we have solved the two-point function of the lowest dimensional scalar operator in the critical $\phi^4$ theory on the $4-\epsilon$ dimensional real projective space in three different ways. The results are consistent with each other, but each method has its own advantage.

The first method we used is the conventional perturbation theory. At order $\epsilon$, the computation is straightforward and we can compute the CFT data with no difficulty. In particular, in the computation of the $\phi$-$\phi$ two-point function, there is no necessity of the renormalization.  Beyond this order, however, the computation becomes more involved and we have to perform the renormalization in the curved background. Note also that the $SO(d+1)$ conformal symmetry on the $\mathbb{RP}^d$ is not manifest in this approach.

The second method we used is the crosscap bootstrap equation. This employs the conformal symmetry manifestly but does not specify the model. In general, we need the infinite number of primary operators to satisfy the crosscap bootstrap equation, but the special features of the CFT data of the critical $\phi^4$ theory allow the truncation. We then found that the crosscap bootstrap equation possesses a one-parameter family of solutions. This corresponds to the existence of the critical $O(N)$ models that satisfy the same crosscap bootstrap equation. We found that once we fix this parameter, e.g. by specifying the anomalous dimension of the $\phi^2$ operator, the solution is unique and coincide with the perturbative computation.

The third method we used is the Schwinger-Dyson equation combined with the conformal symmetry. This approach allows us to evaluate some of the CFT data at $O(\epsilon^2)$  without much a do about the renormalization. Furthermore, we can specify the coupling constant at the critical value even without computing the renormalization group beta function because we imposed the conformal symmetry. On the other hand, we find that not every CFT data is fixed in this approach. In other words, the two-point function as a solution of the Schwinger-Dyson equation contains an integration constant that we cannot determine from this approach alone.
 
As for a future direction, it is a challenging problem to investigate the higher order in $\epsilon$ expansions. The crosscap bootstrap equation must contain infinite number of primary operators. This is because the anomalous dimension $\gamma_{\phi}$ at $O(\epsilon^2)$ is non-zero. Thus, to make progress in analytic approach, we need a certain organizing principle or a resummation to deal with it. In the flat space-time, such techniques have been developed by using the Mellin space formalism in \cite{Gopakumar:2016wkt}\cite{Gopakumar:2016cpb}\cite{Dey:2016mcs}\cite{Dey:2017fab}\cite{Dey:2017oim} as well as in the large spin perturbation theory in \cite{Alday:2017zzv}. We would like to see if a similar technique can be applied to the CFTs on $\mathbb{RP}^d$ or on more non-trivial manifold.

\section*{Acknowledgements}
C.~H. thanks Michio Jimbo for discussions.
C.~H. also thanks the Yukawa Institute for Theoretical Physics at Kyoto University. C.~H. acknowledges discussions during the YITP workshop YITP-W-18-08 on ``Strings and Fields 2017'', which were useful to complete this work.
C.~H. is supported in part by Rikkyo University Special Fund for Research in academic year 2017 (Research category: graduate student package type study, the adoption number 29).
Y.~N. is supported in part by JSPS KAKENHI Grant Number 17K14301.

\section*{Appendices}
\appendix

\section{$\phi$-$\phi$-$\Box^2 \phi^4$ three-point function}\label{phi-phi-Box2 phi4 three point function}
In this appendix, we explicitly compute the three-point function among the two lowest dimensional scalar operators and a primary operator of ``$\Box^{k} \phi^4$''.
Let us focus on the simplest case of $k=2$.
In this case, the explicit form of the primary operator in the schematic notation of ``$\Box^k \phi^4$'' is\footnote{For $k=2,3,4$ the primary operator of ``$\Box^k \phi^4$'', for instance, see subsection 3.2 in \cite{Liendo:2017wsn}.}
\begin{align}
\Box^2 \phi^4 {}_{\mathrm{primary}} := \partial_{\mu} \partial_{\nu} \phi \partial^{\mu} \partial^{\nu} \phi \phi^2 - 4 \partial_{\mu} \partial_{\nu} \phi \partial^{\mu} \phi \partial^{\nu} \phi \phi + 3 \partial_{\mu} \phi \partial_{\nu} \phi  \partial^{\mu} \phi \partial^{\nu} \phi.
\end{align}
The $\phi$-$\phi$-$\Box^2 \phi^4$ three point function can be perturbatively computed as
\begin{align}
\langle \phi (\vec{x}) \phi (\vec{y}) \Box^2 \phi^4 {}_{\mathrm{primary}} (\vec{z}) \rangle^{\mathbb{R}^d} &= \langle \phi (\vec{x}) \phi (\vec{y}) \Box^2 \phi^4 {}_{\mathrm{primary}} (\vec{z}) \rangle_{\mathrm{free}}^{\mathbb{R}^d} \nonumber \\
&\quad - \frac{g}{4!} \int \mathrm{d}^d w \langle \phi (\vec{x}) \phi (\vec{y}) \Box^2 \phi^4 {}_{\mathrm{primary}} (\vec{z}) \phi^4 (\vec{w}) \rangle_{\mathrm{free}}^{\mathbb{R}^d} + O(g^2). \label{eq:phi phi Box2 phi4 3pt func}
\end{align}
As we can see, the first term in \eqref{eq:phi phi Box2 phi4 3pt func} vanishes in the free theory, so the question is if the second term of order $O(g)$ vanishes or not.

To compute the perturbative correction, we use
\begin{align}
\langle \phi (\vec{x}) \phi (\vec{y}) \rangle_{\mathrm{free}}^{\mathbb{R}^d} &= |\vec{x}-\vec{y}|^{-2\Delta_{\phi}^{\mathrm{free}}}, \\
\partial_{\mu} \langle \phi (\vec{x}) \phi (\vec{y}) \rangle_{\mathrm{free}}^{\mathbb{R}^d} &= -2\Delta_{\phi}^{\mathrm{free}} |\vec{x}-\vec{y}|^{-2\Delta_{\phi}^{\mathrm{free}}-2}(x_{\mu}-y_{\mu}), \\
\partial_{\mu} \partial_{\nu} \langle \phi (\vec{x}) \phi (\vec{y}) \rangle_{\mathrm{free}}^{\mathbb{R}^d} &= 4\Delta_{\phi}^{\mathrm{free}} (\Delta_{\phi}^{\mathrm{free}}  + 1) |\vec{x}-\vec{y}|^{-2\Delta_{\phi}^{\mathrm{free}}-4}(x_{\mu}-y_{\mu})(x_{\nu}-y_{\nu}) \nonumber \\ 
&\quad - 2\Delta_{\phi}^{\mathrm{free}} |\vec{x}-\vec{y}|^{-2\Delta_{\phi}^{\mathrm{free}}-2}\delta_{\mu \nu}, 
\end{align}
where $\Delta_{\phi}^{\mathrm{free}}=\frac{d-2}{2}$ is the dimension of the elementary scalar $\phi$ in the free theory.
Note that we denote a $d$-dimensional coordinate vector as $\vec{x}=x^{\mu}, (\mu=1,2,\cdots,d)$. From now on, we set $d=4$ with $\Delta_{\phi}^{\mathrm{free}}=1$.

The $O(g)$ term of \eqref{eq:phi phi Box2 phi4 3pt func} consists of the following three contributions:
\begin{align}
&\langle \phi (\vec{x}) \phi (\vec{y}) \partial_{\mu} \partial_{\nu} \phi \partial^{\mu} \partial^{\nu} \phi \phi^2 (\vec{z}) \phi^4 (\vec{w}) \rangle_{\mathrm{free}}^{\mathbb{R}^d} \nonumber \\
&\quad = \left[ 24 \cdot \langle \phi (\vec{x}) \phi (\vec{y}) \rangle_{\mathrm{free}}^{\mathbb{R}^d} \left[ \partial_{\mu} \partial_{\nu} \langle \phi (\vec{z}) \phi (\vec{w}) \rangle_{\mathrm{free}}^{\mathbb{R}^d} \right] \left[ \partial^{\mu} \partial^{\nu} \langle \phi (\vec{z}) \phi (\vec{w}) \rangle_{\mathrm{free}}^{\mathbb{R}^d} \right] \left[ \langle \phi (\vec{z}) \phi (\vec{w}) \rangle_{\mathrm{free}}^{\mathbb{R}^d} \right]^2 \right. \nonumber \\
&\qquad + 48 \cdot \left[ \partial_{\mu} \partial_{\nu} \langle \phi (\vec{x}) \phi (\vec{z}) \rangle_{\mathrm{free}}^{\mathbb{R}^d} \right] \left[ \partial^{\mu} \partial^{\nu} \langle \phi (\vec{z}) \phi (\vec{w}) \rangle_{\mathrm{free}}^{\mathbb{R}^d} \right] \left[ \langle \phi (\vec{z}) \phi (\vec{w}) \rangle_{\mathrm{free}}^{\mathbb{R}^d} \right]^2 \langle \phi (\vec{w}) \phi (\vec{y}) \rangle_{\mathrm{free}}^{\mathbb{R}^d} \nonumber \\
&\qquad + 48 \cdot \langle \phi (\vec{x}) \phi (\vec{z}) \rangle_{\mathrm{free}}^{\mathbb{R}^d} \left[ \partial_{\mu} \partial_{\nu} \langle \phi (\vec{z}) \phi (\vec{w}) \rangle_{\mathrm{free}}^{\mathbb{R}^d} \right] \left[ \partial^{\mu} \partial^{\nu} \langle \phi (\vec{z}) \phi (\vec{w}) \rangle_{\mathrm{free}}^{\mathbb{R}^d} \right] \langle \phi (\vec{z}) \phi (\vec{w}) \rangle_{\mathrm{free}}^{\mathbb{R}^d} \langle \phi (\vec{w}) \phi (\vec{y}) \rangle_{\mathrm{free}}^{\mathbb{R}^d} \nonumber \\
&\qquad \left. + (\vec{x} \leftrightarrow \vec{y}) \right] \nonumber \\
&\quad = 24 \cdot 48 |\vec{x}-\vec{y}|^{-2} |\vec{z}-\vec{w}|^{-12} \nonumber \\ 
&\qquad + 48 \cdot 64 |\vec{x}-\vec{z}|^{-6} |\vec{z}-\vec{w}|^{-10} (z_{\mu}-x_{\mu})(z^{\mu}-w^{\mu}) (z_{\nu}-x_{\nu})(z^{\nu}-w^{\nu}) |\vec{w}-\vec{y}|^{-2} \nonumber \\
&\qquad - 48 \cdot 16 |\vec{x}-\vec{z}|^{-4} |\vec{z}-\vec{w}|^{-8} |\vec{w}-\vec{y}|^{-2} \nonumber \\
&\qquad + 48 \cdot 48 |\vec{x}-\vec{z}|^{-2} |\vec{z}-\vec{w}|^{-10} |\vec{w}-\vec{y}|^{-2} \nonumber \\
&\qquad + (\vec{x} \leftrightarrow \vec{y}), \label{eq:partial partial phi partial partial phi phi2}
\end{align}
\begin{align}
&-4 \langle \phi (\vec{x}) \phi (\vec{y}) \partial_{\mu} \partial_{\nu} \phi \partial^{\mu} \phi \partial^{\nu} \phi \phi (\vec{z}) \phi^4 (\vec{w}) \rangle_{\mathrm{free}}^{\mathbb{R}^d} \nonumber \\
&\quad = -4 \left[ 24 \cdot \langle \phi (\vec{x}) \phi (\vec{y}) \rangle_{\mathrm{free}}^{\mathbb{R}^d} \left[ \partial_{\mu} \partial_{\nu} \langle \phi (\vec{z}) \phi (\vec{w}) \rangle_{\mathrm{free}}^{\mathbb{R}^d} \right] \left[ \partial^{\mu} \langle \phi (\vec{z}) \phi (\vec{w}) \rangle_{\mathrm{free}}^{\mathbb{R}^d} \right] \left[ \partial^{\nu} \langle \phi (\vec{z}) \phi (\vec{w}) \rangle_{\mathrm{free}}^{\mathbb{R}^d} \right] \langle \phi (\vec{z}) \phi (\vec{w}) \rangle_{\mathrm{free}}^{\mathbb{R}^d} \right. \nonumber \\
&\qquad + 24 \cdot \left[ \partial_{\mu} \partial_{\nu} \langle \phi (\vec{x}) \phi (\vec{z}) \rangle_{\mathrm{free}}^{\mathbb{R}^d} \right] \left[ \partial^{\mu} \langle \phi (\vec{z}) \phi (\vec{w}) \rangle_{\mathrm{free}}^{\mathbb{R}^d} \right] \left[ \partial^{\nu} \langle \phi (\vec{z}) \phi (\vec{w}) \rangle_{\mathrm{free}}^{\mathbb{R}^d} \right] \langle \phi (\vec{z}) \phi (\vec{w}) \rangle_{\mathrm{free}}^{\mathbb{R}^d} \langle \phi (\vec{w}) \phi (\vec{y}) \rangle_{\mathrm{free}}^{\mathbb{R}^d} \nonumber \\
&\qquad + 48 \cdot \left[ \partial_{\mu} \langle \phi (\vec{x}) \phi (\vec{z}) \rangle_{\mathrm{free}}^{\mathbb{R}^d} \right] \left[ \partial_{\nu} \langle \phi (\vec{x}) \phi (\vec{z}) \rangle_{\mathrm{free}}^{\mathbb{R}^d} \right] \left[ \partial^{\mu} \partial^{\nu} \langle \phi (\vec{z}) \phi (\vec{w}) \rangle_{\mathrm{free}}^{\mathbb{R}^d} \right] \langle \phi (\vec{z}) \phi (\vec{w}) \rangle_{\mathrm{free}}^{\mathbb{R}^d} \langle \phi (\vec{w}) \phi (\vec{y}) \rangle_{\mathrm{free}}^{\mathbb{R}^d} \nonumber \\
&\qquad + 24 \cdot \langle \phi (\vec{x}) \phi (\vec{z}) \rangle_{\mathrm{free}}^{\mathbb{R}^d} \left[ \partial_{\mu} \partial_{\nu} \langle \phi (\vec{z}) \phi (\vec{w}) \rangle_{\mathrm{free}}^{\mathbb{R}^d} \right]  \left[ \partial^{\mu} \langle \phi (\vec{z}) \phi (\vec{w}) \rangle_{\mathrm{free}}^{\mathbb{R}^d} \right] \left[ \partial^{\nu} \langle \phi (\vec{z}) \phi (\vec{w}) \rangle_{\mathrm{free}}^{\mathbb{R}^d} \right] \langle \phi (\vec{w}) \phi (\vec{y}) \rangle_{\mathrm{free}}^{\mathbb{R}^d} \nonumber \\
&\qquad \left. + (\vec{x} \leftrightarrow \vec{y}) \right] \nonumber \\
&\quad = - 4 \cdot 24 \cdot 24 |\vec{x}-\vec{y}|^{-2} |\vec{z}-\vec{w}|^{-12} \nonumber \\
&\qquad - 4 \cdot 24 \cdot 32 |\vec{x}-\vec{z}|^{-6} |\vec{z}-\vec{w}|^{-10} (z_{\mu}-x_{\mu})(z^{\mu}-w^{\mu}) (z_{\nu}-x_{\nu})(z^{\nu}-w^{\nu}) |\vec{w}-\vec{y}|^{-2} \nonumber \\
&\qquad + 4 \cdot 24 \cdot 8 |\vec{x}-\vec{z}|^{-4} |\vec{z}-\vec{w}|^{-8} |\vec{w}-\vec{y}|^{-2} \nonumber \\
&\qquad - 4 \cdot 48 \cdot 24 |\vec{x}-\vec{z}|^{-4} |\vec{z}-\vec{w}|^{-10} (z_{\mu}-x_{\mu})(z^{\mu}-w^{\mu}) |\vec{w}-\vec{y}|^{-2} \nonumber \\
&\qquad - 4 \cdot 24 \cdot 24 |\vec{x}-\vec{z}|^{-2} |\vec{z}-\vec{w}|^{-10} |\vec{w}-\vec{y}|^{-2} \nonumber \\
&\qquad + (\vec{x} \leftrightarrow \vec{y}), \label{eq:partial partial phi partial phi partial phi phi}
\end{align}
and
\begin{align}
&3 \langle \phi (\vec{x}) \phi (\vec{y}) \partial_{\mu} \phi \partial_{\nu} \phi  \partial^{\mu} \phi \partial^{\nu} \phi (\vec{z}) \phi^4 (\vec{w}) \rangle_{\mathrm{free}}^{\mathbb{R}^d} \nonumber \\
&\quad = 3 \left[ 24 \cdot \langle \phi (\vec{x}) \phi (\vec{y}) \rangle_{\mathrm{free}}^{\mathbb{R}^d} \left[ \partial_{\mu} \langle \phi (\vec{z}) \phi (\vec{w}) \rangle_{\mathrm{free}}^{\mathbb{R}^d} \right] \left[ \partial^{\mu} \langle \phi (\vec{z}) \phi (\vec{w}) \rangle_{\mathrm{free}}^{\mathbb{R}^d} \right] \left[ \partial_{\nu} \langle \phi (\vec{z}) \phi (\vec{w}) \rangle_{\mathrm{free}}^{\mathbb{R}^d} \right] \left[ \partial^{\nu} \langle \phi (\vec{z}) \phi (\vec{w}) \rangle_{\mathrm{free}}^{\mathbb{R}^d} \right] \right. \nonumber \\
&\qquad \left. + 96 \cdot \left[ \partial_{\mu} \langle \phi (\vec{x}) \phi (\vec{z}) \rangle_{\mathrm{free}}^{\mathbb{R}^d} \right] \left[ \partial^{\mu} \langle \phi (\vec{z}) \phi (\vec{w}) \rangle_{\mathrm{free}}^{\mathbb{R}^d} \right]  \left[ \partial_{\nu} \langle \phi (\vec{z}) \phi (\vec{w}) \rangle_{\mathrm{free}}^{\mathbb{R}^d} \right] \left[ \partial^{\nu} \langle \phi (\vec{z}) \phi (\vec{w}) \rangle_{\mathrm{free}}^{\mathbb{R}^d} \right] \langle \phi (\vec{w}) \phi (\vec{y}) \rangle_{\mathrm{free}}^{\mathbb{R}^d} \right] \nonumber \\
&\qquad \left. + (\vec{x} \leftrightarrow \vec{y}) \right] \nonumber \\
&\quad = 3 \cdot 24 \cdot 16 |\vec{x}-\vec{y}|^{-2} |\vec{z}-\vec{w}|^{-12} \nonumber \\
&\qquad + 3 \cdot 96 \cdot 16 |\vec{x}-\vec{z}|^{-4} |\vec{z}-\vec{w}|^{-10} (z_{\mu}-x_{\mu})(z^{\mu}-w^{\mu}) |\vec{w}-\vec{y}|^{-2} \nonumber \\
&\qquad + (\vec{x} \leftrightarrow \vec{y}). \label{eq:partial phi partial phi partial phi partial phi}
\end{align}
 Since the sum of \eqref{eq:partial partial phi partial partial phi phi2}, \eqref{eq:partial partial phi partial phi partial phi phi}, and \eqref{eq:partial phi partial phi partial phi partial phi} is zero, we find that the $\phi$-$\phi$-$\Box^2 \phi^4$ three point function starts from $O(g^2)$ (i.e. $C_{\phi \phi \Box^2 \phi^4}=O(g^2)=O(\epsilon^2)$). It is interesting to observe that the cancellation happens before the integration over $\vec{w}$.

\section{Formulae for Gauss's hypergeometric function} \label{Formulae for Gauss's hypergeometric function}

In this appendix, we collect the formulae for Gauss's hypergeometric function used in the main text.
Gauss's hypergeometric function has the Taylor expansion:
\begin{align}
{}_{2} F_{1} (a,b;c;\eta) := \sum_{n=0}^{\infty} \frac{(a)_{n}(b)_{n}}{(c)_n n!} \eta^n,
\end{align}
where  $(a)_n$ is the Pochhammer symbol defined by $(a)_0 := 1$ for $n=0$ and $(a)_n := (a)(a+1)\cdots(a+n-1)$ for $n \ge 1$.
For special values of the arguments, we have
\begin{align}
&{}_{2} F_{1} (a,a;a;\eta) = (1-\eta)^{-a}, \\
&\eta^2 \cdot {}_{2} F_{1} (2,2;3;\eta) = 2 \cdot \left[ \ln(1-\eta) + \frac{\eta}{1-\eta} \right].
\end{align}
The following approximation is used in the main text:
\begin{align}
{}_{2} F_{1} (a+\epsilon, a+\epsilon; a+ 2\epsilon; \eta) = (1-\eta)^{-a} + O(\epsilon^2),
\end{align}
based on $(a+\epsilon)_n = (a)_n (1+\epsilon \sum_{k=1}^{n}\frac{1}{a-1+k} + O(\epsilon^2) )$.

\section{Laplacian acting on two-point functions}\label{Laplacian acting two-point functions}
In this appendix, we show the action of the Laplacian on the terms that appear in the conformal partial wave decomposition of the $\mathcal{O}_1$-$\mathcal{O}_2$ two-point function.
In the case of the Laplacian acting once, we find
\begin{align}
& \Box_{x} \left( |\vec{x}-\vec{y}|^{-2\Delta_{\mathcal{O}_1}}  \eta^{\frac{\Delta_{\mathcal{O}_2}}{2} + n} \right) \nonumber \\
&= \left[ (2\Delta_{\mathcal{O}_1} - \Delta_{\mathcal{O}_2} - 2n)(2\Delta_{\mathcal{O}_1} - \Delta_{\mathcal{O}_2} + 2 - d - 2n) \right. \nonumber \\
&\quad - (\Delta_{\mathcal{O}_2} + 2n)(2\Delta_{\mathcal{O}_2} -4\Delta_{\mathcal{O}_1} + 2n)\frac{(\vec{x}-\vec{y}) \cdot \vec{x}}{1+|\vec{x}|^2} \nonumber \\
&\quad \left. - (\Delta_{\mathcal{O}_2} +2n) \left[ (\Delta_{\mathcal{O}_2}+2+2n) - (2 + \Delta_{\mathcal{O}_2} -d)(1+|\vec{x}|^2) \right] \frac{1+|\vec{y}|^2}{1+|\vec{x}|^2} \eta\right] |\vec{x}-\vec{y}|^{-2\Delta_{\mathcal{O}_1}-2}  \eta^{\frac{\Delta_{\mathcal{O}_2}}{2} + n}, \label{eq:Laplacian acting once two-point functions}
\end{align}
where $\eta := \frac{|\vec{x}-\vec{y}|^2}{(1+|\vec{x}|^2)(1+|\vec{y}|^2)}$ is the crosscap crossratio, and $\Delta_{\mathcal{O}_1}$ and $\Delta_{\mathcal{O}_2}$ are the conformal dimension of the local operator $\mathcal{O}_1$ and $\mathcal{O}_2$.

In the case of the Laplacian acting twice, we find
\begin{align}
& \Box_{x} \Box_{y} \left( |\vec{x}-\vec{y}|^{-2\Delta_{\mathcal{O}_1}}  \eta^{\frac{\Delta_{\mathcal{O}_2}}{2} + n} \right) = \left[ a_{(n)}^{\mathcal{O}_1 \mathcal{O}_2} + b_{(n)}^{\mathcal{O}_1 \mathcal{O}_2} \eta + \mathit{O}(\eta^2) \right] |\vec{x}-\vec{y}|^{-2\Delta_{\mathcal{O}_1}-4}  \eta^{\frac{\Delta_{\mathcal{O}_2}}{2} + n}, \label{eq:Laplacian acting twice two-point functions}
\end{align}
\begin{align}
a_{(n)}^{\mathcal{O}_1 \mathcal{O}_2}  &:= ( \Delta_{\mathcal{O}_2} - 2 \Delta_{\mathcal{O}_1} - 2 + 2n )( 2 \Delta_{\mathcal{O}_1} - \Delta_{\mathcal{O}_2} - 2n ) \nonumber \\
& \quad \times ( 2 \Delta_{\mathcal{O}_1} - \Delta_{\mathcal{O}_2} + 2 - d - 2n )( \Delta_{\mathcal{O}_2} - 2 \Delta_{\mathcal{O}_1} - 4 + d + 2n ), \label{eq:coefficient a n O1 O2} \\
b_{(n)}^{\mathcal{O}_1 \mathcal{O}_2} &:= 2( \Delta_{\mathcal{O}_2} + 2n )( 2 \Delta_{\mathcal{O}_1}  - \Delta_{\mathcal{O}_2} - 2n )( 2 \Delta_{\mathcal{O}_1} - \Delta_{\mathcal{O}_2} + 2 - d - 2n )^2  + \mathit{O}(x^2). \label{eq:coefficient b n O1 O2}
\end{align}
Since the above terms \eqref{eq:Laplacian acting once two-point functions} and \eqref{eq:Laplacian acting twice two-point functions} appeared in the expansion \eqref{eq:LHS once Laplacian acting 2pt func for phi} and \eqref{eq:LHS twice Laplacian acting 2pt func for phi}, we need to set a suitable integer number $n$ in order to evaluate the order $\eta$ terms and the order $\eta^2$ terms in the main text.

\section{Laplacian with $\vec{y}=0$} \label{Laplacian with y=0}
In this appendix, we evaluate \eqref{eq: Laplacian acting once 2pt func phi phi in phi4 theory} in the $\vec{y} \to 0$ limit. On the left-hand side, we have
\begin{align}
\lim_{\vec{y} \to 0 }(\mathrm{LHS} \eqref{eq: Laplacian acting once 2pt func phi phi in phi4 theory}) &= \left( \frac{\partial^2}{\partial r^2} + \frac{d-1}{r} \frac{\partial}{\partial r} \right) r^{-2\Delta_{\phi}} \left[ C_{\phi \phi}^{\ \ I}A_{I}^{\pm} \right. \nonumber \\
&\quad + \left[ (C_{\phi \phi}^{\ \ \phi^2}A_{\phi^2}^{\pm})^{(0)} + (C_{\phi \phi}^{\ \ \phi^2}A_{\phi^2}^{\pm})^{(1)}\epsilon \right] \left( \frac{r^2}{1+r^2} \right)^{\frac{\Delta_{\phi^2}}{2}} \sum_{n=0}^{\infty}\frac{\left[ \left(\frac{\Delta_{\phi^2}}{2}\right)_n \right]^2}{(\Delta_{\phi^2}+1-\frac{d}{2})_n n!} \left( \frac{r^2}{1+r^2} \right)^n \nonumber \\
&\quad \left. + (C_{\phi \phi}^{\ \ \phi^4}A_{\phi^4}^{\pm})^{(1)}\epsilon \left( \frac{r^2}{1+r^2} \right)^{\frac{\Delta_{\phi^4}}{2}} \sum_{n=0}^{\infty} \frac{\left[ \left(\frac{\Delta_{\phi^4}}{2}\right)_n \right]^2}{(\Delta_{\phi^4}+1-\frac{d}{2})_n n!} \left( \frac{r^2}{1+r^2} \right)^n + O(\epsilon^2) \right] \nonumber \\
&= C_{\phi \phi}^{\ \ I}A_{I}^{\pm} (2\Delta_{\phi})(2\Delta_{\phi} + 2 - d) r^{-2\Delta_{\phi}-2} \nonumber \\
&\quad + \left[ (C_{\phi \phi}^{\ \ \phi^2}A_{\phi^2}^{\pm})^{(0)} + (C_{\phi \phi}^{\ \ \phi^2}A_{\phi^2}^{\pm})^{(1)}\epsilon \right] \nonumber \\
&\qquad \times \sum_{n=0}^{\infty} \frac{\left[ \left(\frac{\Delta_{\phi^2}}{2}\right)_n \right]^2}{(\Delta_{\phi^2}+1-\frac{d}{2})_n n!} \sum_{m=0}^{\infty} \frac{\left( -\frac{\Delta_{\phi^2}}{2} -n \right)!}{m! \left(-\frac{\Delta_{\phi^2}}{2} -n  -m \right)!} \nonumber \\
&\qquad \times (-2\Delta_{\phi} + \Delta_{\phi^2} + 2n + 2m)(-2\Delta_{\phi} + \Delta_{\phi^2} + 2n + 2m +d -2)r^{-2\Delta_{\phi} + \Delta_{\phi^2} + 2n + 2m -2} \nonumber \\
&\quad + (C_{\phi \phi}^{\ \ \phi^4}A_{\phi^4}^{\pm})^{(1)}\epsilon \sum_{n=0}^{\infty} \frac{\left[ \left(\frac{\Delta_{\phi^4}}{2}\right)_n \right]^2}{(\Delta_{\phi^4}+1-\frac{d}{2})_n n!} \sum_{m=0}^{\infty} \frac{\left( -\frac{\Delta_{\phi^4}}{2} -n \right)!}{m! \left(-\frac{\Delta_{\phi^4}}{2} -n  -m \right)!} \nonumber \\
&\qquad \times (-2\Delta_{\phi} + \Delta_{\phi^4} + 2n + 2m)(-2\Delta_{\phi} + \Delta_{\phi^4} + 2n + 2m +d -2)r^{-2\Delta_{\phi} + \Delta_{\phi^4} + 2n + 2m -2} \nonumber \\
&\quad + O(\epsilon^2)  \nonumber \\
&= 4 C_{\phi \phi}^{\ \ I}A_{I}^{\pm} (\gamma_{\phi})^{(1)}\epsilon r^{-4} \nonumber \\
&\quad + \left[ (C_{\phi \phi}^{\ \ \phi^2}A_{\phi^2}^{\pm})^{(0)} + (C_{\phi \phi}^{\ \ \phi^2}A_{\phi^2}^{\pm})^{(1)}\epsilon \right] \nonumber \\
&\qquad \times \left[ 2\left[ -2(\gamma_{\phi})^{(1)}\epsilon + (\gamma_{\phi^2})^{(1)}\epsilon \right] r^{-2} -8 \left[ 1 -\frac{3}{4}\epsilon + \frac{5}{4} (\gamma_{\phi^2})^{(1)} \epsilon -\frac{3}{2} (\gamma_{\phi})^{(1)}\epsilon \right] \right. \nonumber \\
&\qquad \quad \left. + 8 \left[ 1 -\frac{3}{4}\epsilon + \frac{3}{4} (\gamma_{\phi^2})^{(1)} \epsilon -\frac{3}{2} (\gamma_{\phi})^{(1)}\epsilon \right] \right] \nonumber \\
&\quad + 8 (C_{\phi \phi}^{\ \ \phi^4}A_{\phi^4}^{\pm})^{(1)}\epsilon + O(r^2) + O(\epsilon^2) \nonumber \\
&= 4 C_{\phi \phi}^{\ \ I}A_{I}^{\pm} (\gamma_{\phi})^{(1)}\epsilon r^{-4} \nonumber \\
&\quad + 2 (C_{\phi \phi}^{\ \ \phi^2}A_{\phi^2}^{\pm})^{(0)} \left[ -2(\gamma_{\phi})^{(1)}\epsilon + (\gamma_{\phi^2})^{(1)}\epsilon \right] r^{-2} \nonumber \\
&\quad - 4 (C_{\phi \phi}^{\ \ \phi^2}A_{\phi^2}^{\pm})^{(0)} (\gamma_{\phi^2})^{(1)}\epsilon + 8 (C_{\phi \phi}^{\ \ \phi^4}A_{\phi^4}^{\pm})^{(1)}\epsilon + O(r^2) + O(\epsilon^2).
\end{align}
In the second line, we have used $(1+r^2)^{a} = \sum_{m=0}^{\infty} \frac{a!}{m!(a-m)!} r^{2m}$ with $a = -\frac{\Delta_{\phi^2}}{2}-n$, or  $a = -\frac{\Delta_{\phi^4}}{2}-n$.
On the other hand, the right-hand side of \eqref{eq: Laplacian acting once 2pt func phi phi in phi4 theory} is
\begin{align}
\lim_{\vec{y} \to 0 }(\mathrm{RHS} \eqref{eq: Laplacian acting once 2pt func phi phi in phi4 theory}) &= 
\begin{cases}
\frac{g}{2} (C_{\phi \phi}^{\ \ I}A_{I}^{\pm})^2 (r^{-2}-1+O(r^2)) \\ 
\frac{g}{2} (C_{\phi \phi}^{\ \ I}A_{I}^{\pm})^2 (-r^{-2}-3+O(r^2))
\end{cases}
\end{align}
where $C_{\phi \phi}^{\ \ I}A_{I}^{\pm} = \frac{1}{4\pi^2}$.
Comparing both sides of \eqref{eq: Laplacian acting once 2pt func phi phi in phi4 theory} at $O(\epsilon r^{-4})$, $O(\epsilon r^{-2})$, and $O(\epsilon r^{0})$, we find
\begin{align}
&O(\epsilon r^{-4}): 4 C_{\phi \phi}^{\ \ I}A_{I}^{\pm} (\gamma_{\phi})^{(1)}\epsilon r^{-4}= 0,  \\
&O(\epsilon r^{-2}): 2 (C_{\phi \phi}^{\ \ \phi^2}A_{\phi^2}^{\pm})^{(0)} \left[ -2(\gamma_{\phi})^{(1)}\epsilon + (\gamma_{\phi^2})^{(1)}\epsilon \right] r^{-2} = \frac{g}{2} (C_{\phi \phi}^{\ \ I}A_{I}^{\pm})^2 r^{-2}, \\
&O(\epsilon r^{0}): -4 (C_{\phi \phi}^{\ \ \phi^2}A_{\phi^2}^{\pm})^{(0)} (\gamma_{\phi^2})^{(1)}\epsilon + 8 (C_{\phi \phi}^{\ \ \phi^4}A_{\phi^4}^{\pm})^{(1)}\epsilon =
\begin{cases}
-\frac{g}{2} (C_{\phi \phi}^{\ \ I}A_{I}^{\pm})^2 \\
-\frac{3}{2}g (C_{\phi \phi}^{\ \ I}A_{I}^{\pm})^2,
\end{cases}
\end{align}
where $(C_{\phi \phi}^{\ \ \phi^2}A_{\phi^2}^{\pm})^{(0)} = \pm \frac{1}{4\pi^2} (= \pm C_{\phi \phi}^{\ \ I}A_{I}^{\pm})$ from \eqref{eq:C phi phi phi2 A phi2 in critical phi4 theory}.
Thus, we can reproduce the relations obtained in the main text:
\begin{align}
&(\gamma_{\phi})^{(1)}=0, \label{eq:anomalous dimension phi order epsilon 1 y independent result} \\
&(\gamma_{\phi^2})^{(1)}\epsilon = \frac{g}{16\pi^2}, \label{eq:anomalous dimension phi2 in critical phi4 theory y independent result} \\
&(C_{\phi \phi}^{\ \ \phi^4}A_{\phi^4}^{\pm})^{(1)}\epsilon = \frac{g}{16}(C_{\phi \phi}^{\ \ I}A_{I}^{\pm})^2. \label{eq:C phi phi pih4 A phi4 in critical phi4 theory y independent result}
\end{align}




\end{document}